\documentclass[5p,times,twocolumn]{elsarticle}

\usepackage{graphicx}
\usepackage{amsmath}
\usepackage{amssymb}
\usepackage{booktabs}
\usepackage{array}
\usepackage{multirow}
\usepackage{siunitx}
\DeclareSIUnit{\curie}{Ci}
\DeclareSIUnit{\sps}{S/s}

\usepackage[colorlinks=true,allcolors=blue]{hyperref}
\hypersetup{pdfauthor={The LiquidO Collaboration}}

\journal{Nuclear Instruments and Methods in Physics Research Section A}
\makeatletter
\long\def\MaketitleBox{%
  \resetTitleCounters
  \def\baselinestretch{1}%
  \begin{\elsarticletitlealign}%
   \def\baselinestretch{1}%
    \Large\@title\par\vskip18pt
    \normalsize\elsauthors\par\vskip10pt
    \footnotesize\itshape\elsaddress\par\vskip12pt
  \end{\elsarticletitlealign}}
\renewenvironment{frontmatter}{}{\maketitle
  \twocolumn[{\hrule\vskip12pt
    \ifvoid\absbox\else\unvbox\absbox\par\vskip10pt\fi
    \ifvoid\keybox\else\unvbox\keybox\par\vskip10pt\fi
    \hrule\vskip12pt}]}
\makeatother

\begin{document}

\begin{frontmatter}

\title{Energy, vertex, and time response of a LiquidO detector with an opaque-scintillator target to Compton-tagged electrons}

\author[a]{S.\,Katz}
\author[b]{J.\,T.\,Moffatt}
\author[b]{G.\,Wendel\corref{cor1}}
\cortext[cor1]{\begin{tabular}[t]{@{}l@{}}Corresponding author: \mbox{gmw5164@psu.edu}\\ Collaboration contact: \mbox{LiquidO-Contact-L@in2p3.fr}\end{tabular}}
\author[b,c]{D.\,F.\,Cowen}
\author[a]{I.\,Jovanovic}
\author[d,e]{S.\,Schoppmann}
\author[f]{A.\,S.\,Wilhelm}
\author[g]{J.\,Apilluelo}
\author[h]{L.\,Asquith}
\author[i]{N.\,P.\,Barradas}
\author[j]{J.\,L.\,Beney}
\author[k]{M.\,Berberan~e~Santos}
\author[j]{X.\,de~la~Bernardie}
\author[h]{T.\,J.\,C.\,Bezerra}
\author[j]{M.\,Bongrand}
\author[l]{C.\,Bourgeois}
\author[l]{D.\,Breton}
\author[m]{J.\,Busto}
\author[l,n]{A.\,Cabrera}
\author[j]{A.\,Cadiou}
\author[o]{E.\,Calvo}
\author[h]{M.\,de~Carlos~Generowicz}
\author[h]{B.\,J.\,Cattermole}
\author[p]{E.\,Chauveau}
\author[q]{M.\,Chen}
\author[r]{P.\,Chimenti}
\author[h]{S.\,Kr.\,Das}
\author[s]{S.\,Dusini}
\author[h]{A.\,Earle}
\author[i]{M.\,Felizardo}
\author[r]{C.\,Frigerio~Martins}
\author[g]{J.\,Gal\'an}
\author[g]{J.\,A.\,Garc\'ia}
\author[h]{A.\,Gibson-Foster}
\author[d]{C.\,Girard-Carillo}
\author[t]{J.\,J.\,G\'omez-Cadenas}
\author[h]{W.\,C.\,Griffith}
\author[b]{K.\,Groomes}
\author[j]{M.\,Guiti\`ere}
\author[j]{F.\,Haddad}
\author[h]{J.\,Hartnell}
\author[u]{A.\,Holin}
\author[g]{I.\,G.\,Irastorza}
\author[b]{W.\,Keck}
\author[i]{A.\,Kling}
\author[d]{L.\,Koch}
\author[l,n]{J.\,F.\,Le~Du}
\author[j]{F.\,Lefevre}
\author[l]{P.\,Loaiza}
\author[h]{J.\,A.\,Lock}
\author[v]{T.\,Lux}
\author[g]{G.\,Luz\'on}
\author[l]{J.\,Maalmi}
\author[w]{J.\,P.\,Malhado}
\author[x,y]{F.\,Mantovani}
\author[i]{J.\,G.\,Marques}
\author[p]{C.\,Marquet}
\author[g]{M.\,Mart\'inez}
\author[o]{D.\,Navas-Nicol\'as}
\author[z]{H.\,Nunokawa}
\author[a1]{J.\,P.\,Ochoa-Ricoux}
\author[k]{T.\,Palmeira}
\author[o]{C.\,Palomares}
\author[u]{D.\,Petyt}
\author[j]{P.\,Pillot}
\author[p]{A.\,Pin}
\author[p]{M.\,S.\,Pravikoff}
\author[u]{S.\,Richards}
\author[p]{M.\,Roche}
\author[k]{N.\,Rodrigues}
\author[b1]{R.\,Rosero}
\author[c1]{B.\,Roskovec}
\author[g]{M.\,L.\,Sarsa}
\author[s,d1]{A.\,Serafini}
\author[u]{C.\,Shepherd-Themistocleous}
\author[h]{W.\,Shorrock}
\author[e1]{M.\,Silva}
\author[l]{L.\,Simard}
\author[t]{S.\,R.\,Soleti}
\author[j]{D.\,Stocco}
\author[x,y]{V.\,Strati}
\author[j]{J.\,S.\,Stutzmann}
\author[f1]{F.\,Suekane}
\author[h]{N.\,Tuccori}
\author[o]{A.\,Verdugo}
\author[j]{B.\,Viaud}
\author[d]{S.\,M.\,Wakely}
\author[d]{A.\,Weber}
\author[h,u]{A.\,W.\,R.\,Wong}
\author[b1]{M.\,Yeh}
\author[j]{F.\,Yermia}
\affiliation[a]{organization={Department of Nuclear Engineering, University of California, Berkeley, CA},country={USA}}
\affiliation[b]{organization={Department of Physics, Pennsylvania State University, University Park, PA},country={USA}}
\affiliation[c]{organization={Department of Astronomy and Astrophysics, Pennsylvania State University, University Park, PA},country={USA}}
\affiliation[d]{organization={Johannes Gutenberg-Universit\"{a}t Mainz, Institut f\"{u}r Physik, Mainz},country={Germany}}
\affiliation[e]{organization={Johannes Gutenberg-Universit\"{a}t Mainz, Detektorlabor, Exzellenzcluster PRISMA$^{++}$, Mainz},country={Germany}}
\affiliation[f]{organization={Department of Physics and Nuclear Engineering, United States Military Academy, West Point, NY},country={USA}}
\affiliation[g]{organization={Centro de Astropart\'{\i}culas y F\'{\i}sica de Altas Energ\'{\i}as (CAPA), Universidad de Zaragoza, Zaragoza},country={Spain}}
\affiliation[h]{organization={Department of Physics and Astronomy, University of Sussex, Brighton},country={United Kingdom}}
\affiliation[i]{organization={CT2N, Instituto Superior T\'ecnico, Universidade de Lisboa, Lisbon},country={Portugal}}
\affiliation[j]{organization={Nantes Universit\'e, IMT-Atlantique, CNRS, Subatech, Nantes},country={France}}
\affiliation[k]{organization={iBB, Instituto Superior T\'ecnico, Universidade de Lisboa, Lisbon},country={Portugal}}
\affiliation[l]{organization={Universit\'e Paris-Saclay, CNRS/IN2P3, IJCLab, Orsay},country={France}}
\affiliation[m]{organization={Universit\'e de Aix Marseille, CNRS, CPPM, Marseille},country={France}}
\affiliation[n]{organization={LNCA Underground Laboratory, CNRS, EDF Chooz Nuclear Reactor, Chooz},country={France}}
\affiliation[o]{organization={CIEMAT, Centro de Investigaciones Energ\'{e}ticas, Medioambientales y Tecnol\'{o}gicas, Madrid},country={Spain}}
\affiliation[p]{organization={Universit\'e de Bordeaux, CNRS, LP2I Bordeaux, Gradignan},country={France}}
\affiliation[q]{organization={Department of Physics, Engineering Physics \& Astronomy, Queen's University, Kingston},country={Canada}}
\affiliation[r]{organization={Departamento de F\'isica, Universidade Estadual de Londrina, Londrina},country={Brazil}}
\affiliation[s]{organization={INFN, Sezione di Padova, Padova},country={Italy}}
\affiliation[t]{organization={Donostia International Physics Center, Basque Excellence Research Centre, San Sebasti\'an/Donostia},country={Spain}}
\affiliation[u]{organization={Rutherford Appleton Laboratory, Didcot, Oxford},country={United Kingdom}}
\affiliation[v]{organization={Institut de F\'isica d'Altes Energies (IFAE), The Barcelona Institute of Science and Technology (BIST), Campus UAB, 08193 Bellaterra (Barcelona)},country={Spain}}
\affiliation[w]{organization={Department of Chemistry, Imperial College London, London},country={United Kingdom}}
\affiliation[x]{organization={INFN, Sezione di Ferrara, Ferrara},country={Italy}}
\affiliation[y]{organization={Dipartimento di Fisica e Scienze della Terra, Universit\`{a} di Ferrara, Ferrara},country={Italy}}
\affiliation[z]{organization={Department of Physics, Pontif\'icia Universidade Cat\'olica do Rio de Janeiro, Rio de Janeiro},country={Brazil}}
\affiliation[a1]{organization={Department of Physics and Astronomy, University of California at Irvine, Irvine, CA},country={USA}}
\affiliation[b1]{organization={Brookhaven National Laboratory, Upton, NY},country={USA}}
\affiliation[c1]{organization={Institute of Particle and Nuclear Physics, Charles University, Prague},country={Czech Republic}}
\affiliation[d1]{organization={Dipartimento di Fisica e Astronomia, Universit\`{a} di Padova, Padova},country={Italy}}
\affiliation[e1]{organization={IDMEC, Instituto Superior T\'ecnico, Universidade de Lisboa, Lisbon},country={Portugal}}
\affiliation[f1]{organization={RCNS, Tohoku University, Sendai},country={Japan}}

\begin{abstract}
LiquidO detectors confine scintillation light near its point of production by stochastic scattering in an opaque medium and collect it with a lattice of wavelength-shifting (WLS) fibers.
This configuration gives a monolithic target the imaging power of a segmented detector.
The elementary signal of such a detector is the point-like sub-MeV energy deposition, yet the response to that signal at the center of the detector volume has not been measured against an event-by-event ground truth.
To close this gap, we couple BPULSE (the Berkeley/Penn State Universities LiquidO Science Experiment), a 4-liter prototype LiquidO detector, to a Compton-coincidence calibration setup that selects electron recoils of known energy on a constrained beam axis.
The prototype is a \SI{160}{\milli\meter} cube of opaque scintillator with 64 WLS fibers in alternating orthogonal layers at a \SI{20}{\milli\meter} pitch, read out by silicon photomultipliers.
Compton scatters of collimated gamma-ray beams from $^{137}$Cs and $^{60}$Co sources are tagged by a NaI(Tl) spectrometer.
For \SI{373.4}{keV} electrons, tagged from $90^{\circ}$ scatters of the $^{137}$Cs beam, the detector collects $19.04 \pm 0.07$ photoelectrons (p.e.).
Across four operating points, the two sources at two scattering angles, it responds linearly at $48.3 \pm 1.5$~p.e./MeV\@.
The energy resolution of $(23.4 \pm 0.4)\%$ matches the Poisson statistics of the detected photoelectrons.
Collection at both fiber ends and larger sensors would raise the light yield of this target to an estimated 108--124~p.e./MeV\@.
Center-of-mass reconstruction locates each recoil to the width of the delivered \SI{20}{\milli\meter} beam spot, which bounds the intrinsic localization of the detector from above.
The arrival time response sets a lower bound of \SI{1.5}{\nano\second} on the single-scatter start time at the observed ${\approx}19$~p.e.\ light level.
This single-scatter response is the building block for reconstructing multi-site events.
It constrains the light collection, confinement, and timing models on which reconstruction in future LiquidO detectors will rest.
\end{abstract}


\end{frontmatter}

\section{Introduction}

Proposed programs at the MeV scale, such as the CLOUD reactor-antineutrino experiment~\cite{NavasNicolas:2024ixn} and the NuDoubt++ search for $0\nu\beta^{+}\beta^{+}$~\cite{NuDoubt:2024jax}, require high-fidelity reconstruction of each event's topology to suppress backgrounds.
These experiments must resolve the individual Compton-scatter vertices of the \SI{511}{keV} gamma rays from the annihilation of positrons produced by inverse beta decay or by double-beta-plus decay.
The COCOA Compton telescope extends this requirement to astrophysical MeV gamma rays, imaged by resolving individual Compton vertices inside a LiquidO detector~\cite{LiquidO:2025hbe}.
Such imaging must separate several sub-MeV electron deposits inside one monolithic target and identify the position of each interaction.

This requirement is one motivation for LiquidO~\cite{LiquidO:2019mxd} technology, which has two essential components: a target medium that emits and scatters optical photons when exposed to ionizing radiation, and light guides that capture and transport the light the scattering confines~\cite{Wilhelm:2023hty}.
The spacing of the light guides must match the distance over which the medium confines light, which its scattering length sets.
With the two matched, an opaque target produces event topologies like those of segmented detectors without physical segmentation.

Several LiquidO prototypes have demonstrated the viability and performance of stochastically confined scintillation light collection with WLS fibers over a wide range of energies for electron, muon, and proton signals~\cite{LiquidO:2025qia,LiquidO:2024piw,LiquidO:2025rhe,Che:2026wab}.
No direct measurement of sub-MeV electron recoils of known position and energy at the center of an opaque target volume has been made.
Reconstructing multi-site events, such as the Compton cascades of annihilation gamma rays, starts from the response to a single such recoil.
Measuring it requires a detector volume large enough that light lost at the walls does not distort the response at the center, together with a kinematic constraint on each event.
This constraint cannot easily come from an internal calibration source, as any object placed in the target absorbs a fraction of the photons that would otherwise random walk until capture by a WLS fiber.

We introduce BPULSE (the Berkeley/Penn State Universities LiquidO Science Experiment), a prototype LiquidO detector with a cubic 4-liter opaque volume and an orthogonal lattice of WLS fibers in alternating layers, read out by 64 silicon photomultipliers (SiPMs) with full waveform digitization.
This prototype measures the position, timing, and energy response of an opaque scintillator target to sub-MeV electrons.
Its volume and an external kinematic constraint meet both requirements above.
The constraint comes from a Compton-coincidence technique, in which a collimated gamma-ray beam from an external $^{137}$Cs or $^{60}$Co source enters the target and a NaI(Tl) spectrometer behind an aperture tags the Compton-scattered gamma ray.
The tagged gamma ray fixes the energy of the recoil electron in BPULSE and places its vertex on the beam axis, within the field of view of the aperture.

From single-scatter events at the center of the target we establish the light yield of the opaque medium, its energy resolution, and a vertex localization limited only by the delivered beam spot.
A localization limited by the beam spot means that the measured width is an upper bound on the intrinsic resolution of the detector.
We also characterize the temporal response of the detector to find which component limits its timing.
The emission constants of the scintillator and wavelength shifter dominate it, together with an empirical \SI{2.0}{\nano\second} Gaussian dispersion.
The distance at which an opaque target can separate two energy deposits is set by its scattering length and fiber pitch, both chosen in a range expected for ton-scale LiquidO detectors.
The response reported here constrains the light collection, confinement, and timing models that reconstruction and background suppression in CLOUD and NuDoubt++ will require.
Sections~\ref{sec:apparatus}--\ref{sec:processing} describe the apparatus, its Monte Carlo model, and the selection and photon-counting methods used for these measurements.
Sections~\ref{sec:results} and~\ref{sec:systematics} present the measured response and its uncertainty budget.
Section~\ref{sec:discussion} translates them into design guidance for future opaque detectors.
The appendices describe how measured photoelectrons are extracted from waveforms (\ref{app:rjmcmc}), the selection bias of the fiducial cut, and the dark counts in the time window (\ref{app:selection-derivations}).

\section{The BPULSE Detector and Compton Calibration Setup}

\label{sec:apparatus}

\begin{figure}[htbp]
    \centering
    \includegraphics[width=\columnwidth]{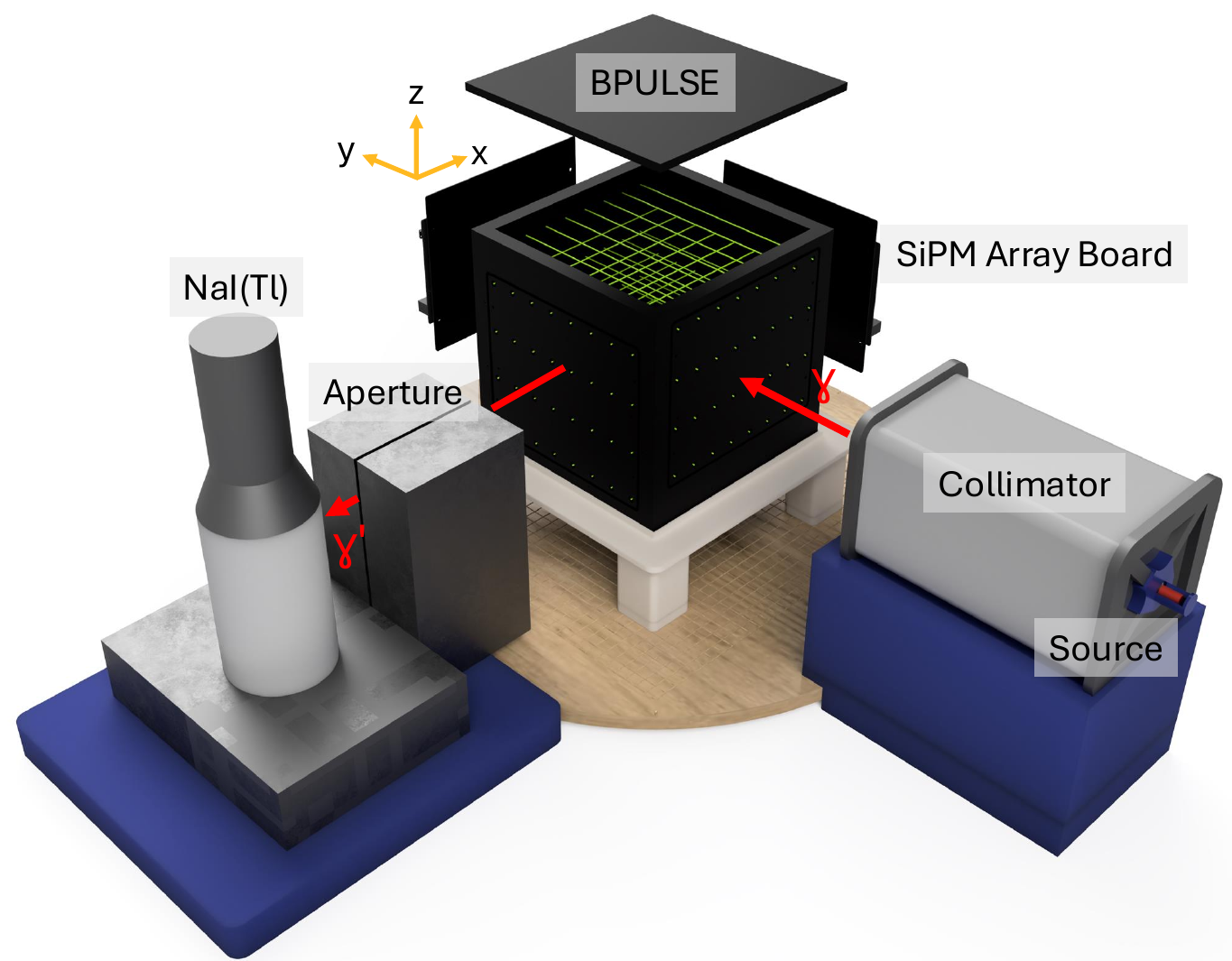}
\caption{Rendering of the BPULSE detector and the Compton calibration setup, exploded to expose the readout: the lid and the two SiPM array boards are displaced from the \SI{160}{\milli\meter} cubic target, revealing the orthogonal WLS-fiber lattice (shown without opaque scintillator).
The dotted faces, opposite the SiPM boards, hold the unread fiber ends, which in the detector are covered by black absorbing caps (not shown).
The collimated gamma-ray beam ($\gamma$) from the $^{137}$Cs or $^{60}$Co source enters along $+Y$.
Gamma rays Compton scattered at $90^{\circ}$ ($\gamma'$) exit along $-X$ through the slit aperture to the NaI(Tl) spectrometer (dimensions in Table~\ref{tab:detector} and Section~\ref{subsec:bench}).}
    \label{fig:apparatus}
\end{figure}

We pair BPULSE with a Compton calibration setup that selects electron recoils of constrained energy and position at the center of the detector.
Figure~\ref{fig:apparatus} shows the setup, in which a collimated gamma-ray beam enters the BPULSE target and gamma rays scattered at the nominal $90^{\circ}$ exit through a slit aperture to the NaI(Tl) tagging spectrometer.
Throughout this paper we use a right-handed coordinate system with the origin at the center of the target volume, $+Y$ along the incident beam, and $+Z$ vertical.
The tagging spectrometer is placed along the $-X$ axis and the scatter occurs in the $(X,Y)$ plane.
The simulation geometry follows this convention.

One readout system records waveforms from both BPULSE and the NaI(Tl) detector, whose trigger is independent of the BPULSE response.
Table~\ref{tab:detector} lists the main detector parameters.

\subsection{The BPULSE Detector}
\label{subsec:detector}

The BPULSE detector is a \SI{160}{\milli\meter} cubic active volume, approximately \SI{4.1}{\liter} of target medium, inside a CNC-machined shell of \SI{10}{\milli\meter} black cast acrylic.
The scintillator is a variant of NoWaSH (New opaque Wax Scintillator, Heidelberg)~\cite{Buck:2019tsa}: purified linear alkylbenzene (LAB) with the fluors 2,5-diphenyloxazole (PPO) and 1,4-bis(2-methylstyryl)benzene (bis-MSB)\@.
Their concentrations are close to those of the JUNO scintillator, \SI{2.5}{\gram\per\liter} PPO and \SI{3}{\milli\gram\per\liter} bis-MSB, whose emission time profile~\cite{Beretta:2025jinst} the timing model of Section~\ref{subsec:res-timing} uses.
It is loaded with approximately 9\% by weight of micronized wax, which is expected to produce millimeter-scale scattering lengths.
At this loading the mixture does not fully solidify, and a bulk of saturated wax forms beneath a ${\approx}\SI{10}{\milli\meter}$ layer of clear LAB at the top of the volume.
The wax loading was kept slightly below the solidification threshold to avoid the stress cracks observed at higher loadings and to limit absorption by impurities in the unpurified wax.
At the \SI{20}{\milli\meter} pitch this absorption couples strongly to the light collection; it is part of the measured light yield and cannot be separated from the scattering length (Section~\ref{subsec:res-timing}).
The scattering and absorption lengths of this mixture have not been measured, so they are treated as order-of-magnitude estimates (Table~\ref{tab:detector}).

The active volume is instrumented with 64 multi-clad Kuraray Y-11(200) WLS fibers of \SI{1}{\milli\meter} diameter in two orthogonal directions.
The fibers are arranged in eight horizontal layers of eight parallel fibers, alternating between $X$ and $Y$ orientation (the lowest layer along $X$) at a \SI{20}{\milli\meter} pitch within each layer and a \SI{20}{\milli\meter} layer spacing along $Z$.
Each layer gives one transverse coordinate, $X$ or $Y$, with the layer number giving $Z$; the lattice thus records a stereo set of 2D projections of the interaction volume.

The fibers pass through a black acrylic shell and are attached with epoxy, cut flush, and mechanically polished flat to the acrylic surface.
One end of each fiber is read out by a SiPM (Section~\ref{subsec:daq}); the other end is not read out and ends in air under an optically black 3D-printed cap.
A reflecting far end would return light with an uncalibrated reflectivity and after a second transit of the fiber, which smears the timing response.
The absorbing end removes this unknown reflectivity from the light yield of every channel, at a cost of a factor of two in collected light, as quantified in Section~\ref{sec:discussion}.

\begin{table}[t]
\centering
\small
\caption{Detector specifications.}
\label{tab:detector}
\begin{tabular}{@{}l >{\raggedright\arraybackslash}p{5.6cm}@{}}
\toprule
\multicolumn{2}{@{}l}{\textit{Vessel and scintillator}} \\
Active volume & $(\SI{160}{\milli\meter})^3$ (${\approx}\SI{4.1}{\liter}$) \\
Shell & \SI{10}{\milli\meter} black cast acrylic \\
Formulation & NoWaSH: LAB + ${\approx}9$\,wt\% micronized wax; PPO ${\approx}\SI{2.5}{\gram\per\liter}$, bis-MSB ${\approx}\SI{3}{\milli\gram\per\liter}$ \\
Emission peak & ${\approx}\SI{430}{\nano\meter}$ (bis-MSB) \\
Scattering & mm-scale (wax loading); unmeasured \\
Absorption & unmeasured \\
\midrule
\multicolumn{2}{@{}l}{\textit{WLS fiber lattice}} \\
Fibers & 64 Kuraray Y-11(200), \SI{1}{\milli\meter} diameter, multi-clad; absorption peak ${\approx}\SI{430}{\nano\meter}$ \\
Geometry & 8 layers $\times$ 8, alternating $X$/$Y$, \SI{20}{\milli\meter} pitch \\
\midrule
\multicolumn{2}{@{}l}{\textit{Readout}} \\
Photosensors & Hamamatsu S13360-1350PE ($1.3 \times 1.3~\mathrm{mm}^2$), \SI{+3}{\volt} overvoltage, \SIrange{31}{33}{\degreeCelsius} \\
Digitizer & Nalu HDSoCv2, \SI{1}{\giga\sps}, \SI{864}{\nano\second} record; 60 SiPM + 1 NaI(Tl) channels \\
\midrule
\multicolumn{2}{@{}l}{\textit{Compton calibration setup}} \\
Collimator & \SI{180}{\milli\meter} lead, \SI{9.25}{\milli\meter} bore; beam diameter ${\approx}\SI{20}{\milli\meter}$ at the target center \\
Aperture & \SI{100}{\milli\meter} lead, \SI{3}{\milli\meter} nominal slit (3--\SI{4}{\milli\meter} measured) \\
Crystal & $3'' \times 5''$ NaI(Tl) (Bicron 3M5/3-X) \\
\bottomrule
\end{tabular}
\end{table}

\subsection{The Compton Calibration Setup and Tagging Kinematics}
\label{subsec:bench}

Interchangeable isotropic $^{137}$Cs (\SI{481}{\mega\becquerel}, \SI{13}{\milli\curie}) and $^{60}$Co (\SI{1.44}{\giga\becquerel}, \SI{39}{\milli\curie}) sources are collimated into the primary gamma-ray beam.
Each source is measured at scattering angles of $45^{\circ}$ and $90^{\circ}$, and each source and angle pair is an operating point.
A fixed holder keeps each source in the same position, and only the NaI(Tl) spectrometer moves between the two angles.
The nominal operating point is $^{137}$Cs at $90^{\circ}$, because $^{137}$Cs emits a single line of known energy per decay and a $90^{\circ}$ scatter gives a recoil region close to a right cylinder (Section~\ref{subsec:expected-signal}).
The vertex, timing, resolution, and sensor results are quoted at the nominal point.
The other three operating points enter the light yield linearity (Section~\ref{subsec:res-lightyield}).
There the fiducial cut, a bound on the reconstructed vertex around the volume center (Section~\ref{subsubsec:fiducial}), removes the geometry differences between oblique and right-cylinder recoil regions.
The source is located in a lead cave ${\approx}\SI{410}{\milli\meter}$ from the target center, behind a \SI{180}{\milli\meter} lead collimator ($96.8 \times 96.8~\mathrm{mm}^2$ cross section, cast in a stainless-steel casing) with a \SI{9.25}{\milli\meter} steel-lined bore that defines the beam.
With the exit of the collimator ${\approx}\SI{210}{\milli\meter}$ from the origin, the beam is approximately \SI{20}{\milli\meter} in diameter at the center of the active volume.

Gamma rays that Compton scatter within the BPULSE target at the selected angle are detected by the cylindrical NaI(Tl) tagging spectrometer (Bicron 3M5/3-X, \SI{38.1}{\milli\meter} radius, \SI{127.0}{\milli\meter} length).
The spectrometer is mounted with its axis vertical and its center approximately \SI{340}{\milli\meter} from the detector origin.
A CNC-machined dial with precise angular markings selects the scattering angle.
It couples to interlocking 3D-printed alignment blocks that set the spacing of the lead shielding and orient the source collimator and aperture relative to the BPULSE target.

The \SI{100}{\milli\meter}-thick lead aperture with its \SI{3}{\milli\meter} vertical slit restricts the field of view of the tagging spectrometer, constraining the scattering angle $\phi$ between the incident and scattered gamma-ray directions.
The measured slit width varies between 3 and \SI{4}{\milli\meter} due to lead brick imperfections and alignment.
The width of the tagged recoil energy is bounded by this constraint and by the beam width.
A ray trace of the aperture, combined with the bound-electron Doppler term~\cite{Ribberfors:1975}, gives an intrinsic width of $\sigma \approx \SI{4}{keV}$ for the tagged line, about $1\%$ of the recoil energy and negligible relative to the detector resolution.\label{subsubsec:width-budget}
With the 3--\SI{4}{\milli\meter} slit variation taken into account, the intrinsic width of the tagged line increases by ${\lesssim}10\%$.
For a given incident gamma-ray energy $E_{\gamma}$ and scattering angle $\phi$, the scattered gamma-ray energy $E_{\gamma}'$ is uniquely determined by the Compton scattering formula
\begin{equation}
E_{\gamma}' = \frac{E_{\gamma}}{1 + \frac{E_{\gamma}}{m_e c^2}(1 - \cos\phi)}.
\label{eq:compton}
\end{equation}

The recoil electron carries the remaining energy, $E_e = E_{\gamma} - E_{\gamma}'$, and the tagged photopeak in the spectrometer provides the event ground truth for the offline selection.
The measured operating points are listed in Table~\ref{tab:kinematics}.
At the nominal operating point ($^{137}$Cs at $90^{\circ}$) the NaI(Tl) detector selects $E_{\gamma}' = \SI{288.3}{keV}$ gamma rays associated with $E_{e^-} = \SI{373.4}{keV}$ recoils in BPULSE\@.
$^{60}$Co lines are tagged together as an unresolved doublet.
The Klein--Nishina~\cite{KleinNishina:1929} weights $f_{\mathrm{KN}}$ of the two lines (Table~\ref{tab:kinematics}), with the NaI window acceptance of each run included, set the mean recoil energy at each angle.
Effective recoil energies are ${\approx}\SI{889}{keV}$ at $90^{\circ}$ and ${\approx}\SI{523}{keV}$ at $45^{\circ}$, where the energy windows accept the two $^{60}$Co lines with nearly equal efficiency.

\begin{table}[t]
\centering
\small
\caption{Operating points measured with the Compton calibration setup; energies in keV\@.
$f_{\mathrm{KN}}$ is the acceptance-corrected Klein--Nishina weight of each line within the unresolved $^{60}$Co doublet. $N_{\mathrm{sel}}$ is the number of events passing the selection of Section~\ref{subsec:selection} and $N_{\mathrm{fid}}$ the number of those in the fiducial volume.}
\label{tab:kinematics}
\begin{tabular}{@{}lcccccrr@{}}
\toprule
Source & $\phi$ & $E_{\gamma}$ & $E_{\gamma}'$ & $E_e$ & $f_{\mathrm{KN}}$ & $N_{\mathrm{sel}}$ & $N_{\mathrm{fid}}$ \\
\midrule
$^{137}$Cs & $45^{\circ}$ & 661.7  & 479.7 & 181.9 & 1 & $143{,}957$ & $2036$ \\
$^{137}$Cs & $90^{\circ}$ & 661.7  & 288.3 & 373.4 & 1 & $139{,}527$ & $3804$ \\
\midrule
\multirow{2}{*}{$^{60}$Co} & \multirow{2}{*}{$45^{\circ}$} & 1173.2 & 701.5 & 471.7 & 0.51 & \multirow{2}{*}{$386{,}079$} & \multirow{2}{*}{$8628$} \\
                           &                              & 1332.5 & 755.5 & 577.0 & 0.49 & & \\
\addlinespace
\multirow{2}{*}{$^{60}$Co} & \multirow{2}{*}{$90^{\circ}$} & 1173.2 & 356.0 & 817.2 & 0.51 & \multirow{2}{*}{$132{,}924$} & \multirow{2}{*}{$4001$} \\
                           &                              & 1332.5 & 369.4 & 963.1 & 0.49 & & \\
\bottomrule
\end{tabular}
\end{table}

\subsection{Readout Electronics, Data Acquisition, and Trigger}
\label{subsec:daq}

Photons from WLS fibers are detected by Hamamatsu S13360-1350PE SiPMs ($1.3 \times 1.3~\mathrm{mm}^2$ active area) at the detector boundaries.
Each SiPM is operated at \SI{+3}{\volt} above the breakdown voltage that the manufacturer measured for it.
It is coupled to the polished fiber end through ${\approx}\SI{0.2}{\milli\meter}$ of optical grease (EJ-550 optical-grade silicone, refractive index 1.46) and a \SI{0.3}{\milli\meter} epoxy window (refractive index 1.55), about \SI{0.5}{\milli\meter} from fiber face to silicon.
The system is cooled by forced air.
The sensors ran at \SIrange{31}{33}{\degreeCelsius} during the campaign, and TMP119 temperature sensors adjust the per-channel bias continuously to compensate for thermal drift.
The residual drift is negligible compared to the channel-to-channel variation removed by the per-channel calibration.
The channel-to-channel variation in gain is measured \textit{in situ} at $4\%$ (Section~\ref{subsec:spe}).
The SiPM pulses are shaped by OPA847 transimpedance amplifiers.

A single 64-channel HDSoCv2 digitizer~\cite{Macchiarulo:2023hdsocv2,Macchiarulo:2024hdsocv2} records the response of both BPULSE and the NaI(Tl) detector at a sampling rate of \SI{1}{\giga\sps}.
The HDSoCv2 is a system-on-chip with a switched capacitor array that records 28 blocks of 32 samples on each trigger.
The first 32-sample block of each record is discarded because its baseline carries an unstable \SI{+39}{\milli\volt} offset, leaving an \SI{864}{\nano\second} usable record per channel.
The hardware trigger typically places the tag pulse \SIrange{50}{180}{\nano\second} into this usable record.
Array light from BPULSE precedes it by the ${\approx}\SI{42}{\nano\second}$ NaI(Tl) transit and cable offset (Fig.~\ref{fig:selection}a).
One of the 64 SiPM channels is used to digitize the NaI(Tl) tagging spectrometer, leaving one fiber on the $Y$-normal beam entrance wall without readout.
Three channels on the chip had an offset that changed from one 32-sample block to the next and were deemed unreliable for single photon counting.
With these excluded, 60 live SiPM channels are available for analysis.
Figure~\ref{fig:faces} shows this channel and the three excluded channels in black, all on the $Y$-normal beam entrance wall.

The tagging spectrometer generates the hardware trigger: its \SI{50}{\ohm} transmission line is split by a high-impedance tap to an oscilloscope and a logic-gate pulse shaper.
A runt discriminator trigger requires a peak amplitude between two thresholds and propagates the signal on the falling edge of the lower-threshold crossing.
The two thresholds define a hardware energy window on the scattered gamma-ray energy before any software selection is applied.
Accepted gamma rays span \SIrange{161}{386}{keV} for the nominal configuration via the calibration given in Section~\ref{subsec:lowlevel}.
The window is wider than the tagged line, so that the software cut on the tag amplitude (Section~\ref{subsec:selection}) sets the final energy selection.
The thresholds are reset at each operating point so that the window contains the corresponding scattered line (plateaus \SIrange{228}{481}{keV} at $^{60}$Co $90^{\circ}$, \SIrange{318}{643}{keV} at $^{137}$Cs $45^{\circ}$, and \SIrange{548}{893}{keV} at $^{60}$Co $45^{\circ}$).

Data at each operating point were acquired in a single continuous run, \SI{85.7}{\hour} at a mean accepted trigger rate of \SI{4.3}{\hertz} for the nominal $^{137}$Cs $90^{\circ}$ configuration ($1.32\times10^{6}$ triggers).
Three other operating points ran \SI{26.9}{\hour} at \SI{19.8}{\hertz} ($^{60}$Co $90^{\circ}$, $1.92\times10^{6}$ triggers), \SI{20.4}{\hour} at \SI{17.8}{\hertz} ($^{60}$Co $45^{\circ}$, $1.31\times10^{6}$ triggers), and \SI{15.8}{\hour} at \SI{6.7}{\hertz} ($^{137}$Cs $45^{\circ}$, $3.8\times10^{5}$ triggers).

\section{Monte Carlo Simulation and Expected Signal Characteristics}
\label{sec:simulation}

A Monte Carlo simulation of the entire setup was implemented in the Geant4-based~\cite{GEANT4:2002zbu,Allison:2006ve,Allison:2016lfl} RATPAC-two~\cite{ratpac-two} framework and split into two stages, because optical photon transport dominates the computational cost per event.
In the first stage, gamma-ray transport alone gives the expected distribution of electron deposits in BPULSE\@. This distribution sets the size of the tagged interaction region of interest and the frequency of degenerate multi-scatter recoils. In the second stage, a subset of these events is passed to the optical model.

\subsection{Transport Simulation and the Tagged-Electron Ground Truth}
\label{subsec:expected-signal}

The simulation reproduces the geometry of the calibration setup and assigns the BPULSE active volume the NoWaSH composition of Table~\ref{tab:detector}.
Gamma rays are emitted into an $8^{\circ}$ half-angle cone that covers the collimator bore.
When the energy deposited in the tagging crystal is within ${\approx}\pm\SI{100}{keV}$ of the expected scattered gamma-ray line, the event and its global interaction history are recorded.
The $^{137}$Cs productions at $45^{\circ}$ and $90^{\circ}$ were each run at both slit widths (3 and \SI{4}{\milli\meter}), with ${\approx}2\times10^{13}$ source decays each.
Values in this section refer to the nominal \SI{3}{\milli\meter} slit, with the \SI{4}{\milli\meter} production given where it differs.

\label{subsubsec:multiscatter}A gamma ray that reaches the tagging crystal has interacted in BPULSE zero, one, or several times.
Gamma rays that interact only in the tagging crystal add to the flat accidental background, which is subtracted statistically in the light yield analysis.
A single scatter in the field of view of the aperture places the event in the tag photopeak, its energy fixed by two-body kinematics.
Several consecutive shallow scatters can deposit the correct energy in BPULSE and in the tagging detector.
These multi-scatter histories are 24\% of tagged interactions with tag energy within \SI{25}{keV} of the nominal \SI{288.3}{keV} line.
Only the signal in BPULSE separates them from the single scatters.

The region of interest is defined by the spatial distribution of single-scatter events, the signal sample.
From a density optimization on the simulated interaction vertices, this region is a cylinder of radius $R \le \SI{12}{\milli\meter}$ around the beam axis, with an axial extent defined by the slit.
Ray traces at 3 and \SI{4}{\milli\meter} slit widths give an axial extent of $|Y| \le 7.1$ and \SI{9.3}{\milli\meter}, respectively.
Within the energy window, this region contains $89.5\%$ of tagged single scatters at a $(92.6 \pm 0.3)\%$ single-scatter fraction.

\begin{figure}[t]
    \centering
    \includegraphics[width=\columnwidth]{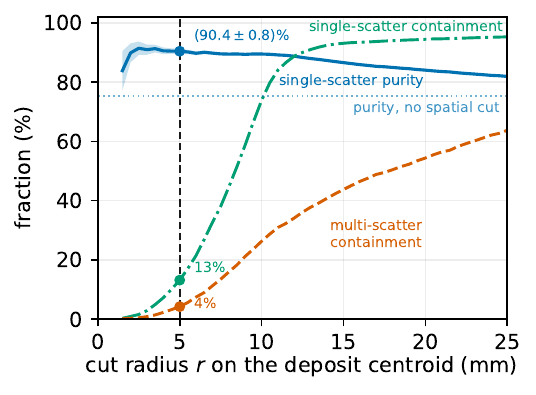}
\caption{Geant4 transport simulation: single-scatter purity and containment as a function of the radial cut on the energy-weighted centroid ($90^{\circ}$ simulation, \SI{3}{\milli\meter} aperture).
The centroid is the truth-level analogue of the reconstructed center-of-mass vertex.
Markers and the dashed line: the \SI{5}{\milli\meter} fiducial volume adopted in the event selection (band: binomial uncertainty).}
    \label{fig:roi-purity}
\end{figure}

\label{subsubsec:fiducial}In the data, where the true vertex is unavailable, the fiducial volume is defined on the center-of-mass vertex of Section~\ref{subsec:observables}, which estimates the true energy-weighted centroid from the detected light.
In the simulation, the median distance of the true centroid from the volume center is \SI{8.4}{\milli\meter} for tagged single scatters and \SI{17.7}{\milli\meter} for multi-scatter events (the $50\%$ crossings shown in Fig.~\ref{fig:roi-purity}).
A \SI{5}{\milli\meter} bound on the true centroid gives $(90.4 \pm 0.8)\%$ single-scatter purity and keeps $13\%$ of the tagged single scatters (the signal efficiency), while accepting $4.2\%$ of the multi-scatter events.
On the reconstructed vertex the same bound keeps $3$--$4\%$ of the simulated single scatters, consistent with the $2.7\%$ of the selected data sample that it keeps (Table~\ref{tab:kinematics}).
With the \SI{4}{\milli\meter} aperture the purity is $(89.3 \pm 0.7)\%$ at $11\%$.

\subsection{Optical and Sensor Model}
\label{subsec:optical-model}

The optical model supplies the center-of-mass vertex benchmark, the transport kernels of the timing analysis, and the capture ratios of the sensor coupling.
For each electron passed from the transport stage, it follows the light from scintillation to the SiPM photocathode.
The opacity of the scintillator is modeled as a spherically symmetric Rayleigh-type scattering process with a scattering length decoupled from the bulk absorption length, making the parameter an isotropic-equivalent transport length.
For the fiber core, the simulation uses the Y-11 absorption spectrum of Refs.~\cite{Wilhelm:2023hty,Herzkamp2016SimulationNeutrons}, whose absorption length is sub-millimeter across the ${\approx}395$--\SI{470}{\nano\meter} band and rises to meters above ${\approx}\SI{480}{\nano\meter}$.
Photons absorbed in the core are re-emitted with unit probability on the Y-11 emission spectrum given in the same references, which peaks near \SI{478}{\nano\meter}.
The attenuation along the fiber is modeled, but over the ${\approx}\SI{10}{\centi\meter}$ photon paths in these fibers, the attenuation is negligible compared to the meter-scale attenuation length.
Every material interface uses the Geant4 \textsc{unified} surface model~\cite{Levin:1996}, with all finishes polished.
The refractive indices are 1.59 for the fiber core and 1.49 and 1.42 for the claddings.
In the vertex and timing simulations each fiber end faces its sensor across a \SI{0.5}{\milli\meter} air gap and a glass window of refractive index 1.46.
The coupling therefore enters those results only as an overall collection efficiency, which the scaling to the measured light level removes.
The sensor-coupling study of Section~\ref{sec:discussion} models the grease and epoxy window of Section~\ref{subsec:daq}.
A set of photon arrival times is produced at each sensor.
Dark counts are added in post-processing at the rate recorded in a background run. The analog chain is modeled outside of RATPAC.

Predictions of absolute light collection are not used because coefficients that set the collection scale are unmeasured and mutually degenerate: the absorption and scattering lengths of the scintillator, the quantum yield of the WLS dye, and imperfections in fiber/coupling.
The data constrain only their combination; separating them is left to future studies.
The center-of-mass reconstruction study alone required the absolute light collection to match the data.
In that study, the simulation was run at an overestimated light yield and thinned by binomial sampling to match the light yield in data.

\section{Signal Processing, Event Selection, and Photon Counting}
\label{sec:processing}

The hardware trigger of Section~\ref{subsec:daq} decides only which events are recorded.
All later processing is performed in software.
Feature extraction first reduces each NaI(Tl) and BPULSE waveform to a handful of per-channel numbers: charge, amplitude, and arrival time.
These scalars are assigned to the tagged coincidence events of Fig.~\ref{fig:selection} and Table~\ref{tab:selection}.
While these quantities are sufficient for basic event selection, they contain excess noise and nonlinearities.
A single photon characterization corrects the systematics of these scalars and builds the forward model that photon counting requires.
With this model, the number and arrival time of every individual photon in each waveform are reconstructed.
A second selection, with a cut on the reconstructed vertex, follows from these counts.

\subsection{Waveform Feature Extraction}
\label{subsec:lowlevel}

The energy and arrival time of each NaI(Tl) pulse are measured from the waveform with an empirical template, the average NaI(Tl) pulse shape measured in data, chosen because it is insensitive to clipping of pulse tails in the recorded data.
This noise-whitened template, $\mathcal{T}_{\mathrm{NaI}}(t)$, is built from data taken with calibration sources ($^{137}$Cs, $^{60}$Co, and $^{22}$Na).
A linear least-squares fit of the scaled template to each baseline-subtracted waveform yields the sub-sample arrival time and the amplitude $\hat{A}_{\mathrm{NaI}}$.
For each trial time shift the best amplitude follows in closed form, so the fit searches over the time shift alone (Brent's method~\cite{Brent:1973}).
A constant-fraction-discriminator (CFD) estimate starts this search.
The template is shifted by fractions of a sample in the frequency domain.
The fitted amplitude converts to energy through a linear fit to these check-source photopeaks (\SIrange{511}{1332.5}{keV}):
\begin{equation}
E_{\gamma}'\,[\mathrm{keV}] = 2.578\,\hat{A}_{\mathrm{NaI}}\,[\mathrm{mV}] - 25.9.
\label{eq:nai-calibration}
\end{equation}
Fitted $\hat{A}_{\mathrm{NaI}}$ and $t_{\mathrm{NaI}}$ are inputs to the event selection.

\begin{figure*}[t]
    \centering
    \includegraphics[width=\textwidth]{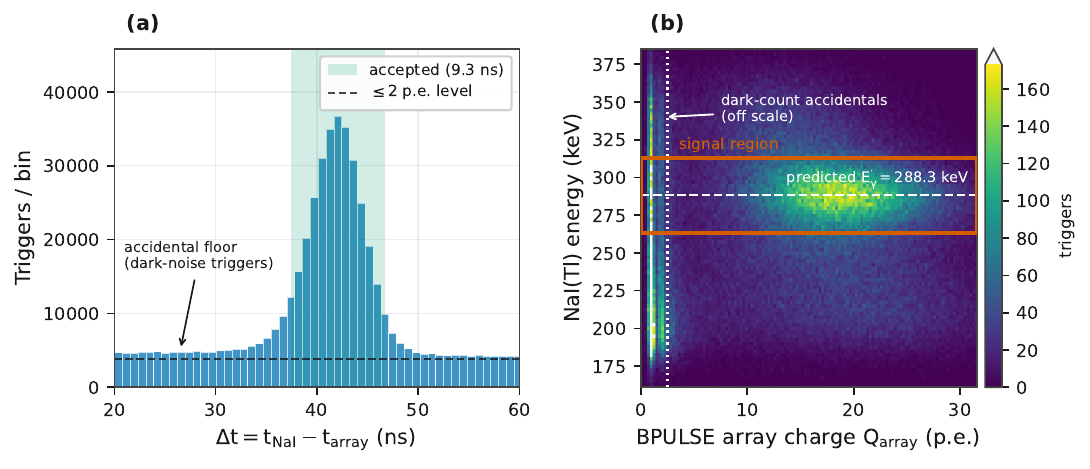}
\caption{Kinematic event selection (data, $^{137}$Cs); each panel shows one selection variable with the complementary selection applied.
(a) Coincidence time $\Delta t = t_{\mathrm{NaI}} - t_{\mathrm{array}}$ with the coincidence window shaded. The window is $\pm 1.8$ standard deviations of the fitted coincidence peak ($\sigma \approx \SI{2.6}{\nano\second}$).
The flat background is the dark-noise accidental population, its ${\leq}2$~p.e.\ component dashed.
Beyond ${\approx}\SI{60}{\nano\second}$ the early edge of the time window truncates the distribution.
(b) NaI(Tl) energy (Eq.~\eqref{eq:nai-calibration}) as a function of array charge $Q_{\mathrm{array}}$ inside the coincidence window, in single-p.e.\ units, with the signal region outlined and the predicted $E_{\gamma}'$ (Table~\ref{tab:kinematics}) dashed; the color scale saturates at the tagged-recoil peak, and the accidental band left of the dotted 2~p.e.\ line runs off scale.}
    \label{fig:selection}
\end{figure*}

Each BPULSE SiPM waveform is passed through a low-pass filter and compared to a discrimination threshold.
Waveforms above this threshold register a hit and give the charge over threshold $Q$.
A constant-fraction estimate of the hit arrival time then tests whether it falls in the time window around the tag defined in Section~\ref{subsec:observables}.
The accepted hits then give two quantities for the array: the total charge $Q_{\mathrm{array}}$ and the global interaction time $t_{\mathrm{array}}$.
The latter is the time at which the summed waveform of the contributing channels reaches $20\%$ of its peak.
Uncorrected timing offsets between channels, \SI{0.2}{\nano\second} RMS from trace lengths and a coherent \SI{0.8}{\nano\second} between the two 32-channel halves of the HDSoCv2 digitizer, are negligible compared to the coincidence width (Fig.~\ref{fig:selection}a).

\subsection{Event Selection}
\label{subsec:selection}

\begin{table}[t]
\centering
\caption{The five event populations after selection and their share of the selected sample ($139{,}527$ events), at the nominal $^{137}$Cs $90^{\circ}$ operating point.
The coincident $^{60}$Co cascade gamma is not populated here.}
\label{tab:selection}
\footnotesize
\setlength{\tabcolsep}{3pt}
\begin{tabular}{@{}lrl@{}}
\toprule
Population & Share & Source \\
\midrule
Single-scatter signal & $(68.7 \pm 0.5)\%$ & Simulation \\
Multi-scatter background & $(22.7 \pm 0.4)\%$ & Simulation \\
Accidental coincidences & $(8.6 \pm 0.4)\%$ & data \\
Cosmic rays & $<0.1\%$ & data \\
Coincident $^{60}$Co cascade gamma & --- & --- \\
\bottomrule
\end{tabular}
\end{table}

The hardware trigger accepts every NaI(Tl) pulse in the trigger energy window, so the recorded sample mixes the single-scatter recoils with four other populations (Table~\ref{tab:selection}): accidental coincidences of an unrelated tag with unrelated BPULSE light, multiple scatters in the BPULSE volume, cosmic rays through both detectors, and, at the $^{60}$Co points, the coincident cascade gamma from the same decay.
Three cuts on the extracted scalars, on the coincidence time, the tag amplitude and the BPULSE charge, address the accidentals, the recoil energy and the cosmic rays.
The multi-scatter background is handled outside these cuts.

Accidental coincidences arrive uniformly in time.
A cut at 1.8 standard deviations of the fitted NaI(Tl)--BPULSE coincidence peak removes most of them (Fig.~\ref{fig:selection}).
The remainder, $8.6\%$ of the selected sample (Table~\ref{tab:selection}) at a constant rate in time, is subtracted statistically.
The window width lies on a plateau where widening it no longer changes the subtracted light yield.
Widening it to $\pm\SI{14}{\nano\second}$ changes the central mean by $<0.4\%$.

The tag amplitude cut, 1.5 standard deviations about the fitted photopeak, fixes the recoil energy.
It accepts $87\%$ of the photopeak coincidences, but photopeak events are only $72\%$ of the accepted sample after accidental subtraction.
The rest are mostly multiple scatters in the BPULSE volume.
They deposit the same total energy as a single recoil, so the energy deposited in BPULSE cannot separate them.
Their tag energies form a continuum, nearly flat across the photopeak, that no window removes.
They are $(24.8 \pm 0.4)\%$ of true coincidences in the transport simulation and $(25 \pm 4)\%$ in the fit of the pedestal under the peak in data; the $22.7\%$ of Table~\ref{tab:selection} is the same population over the full selected sample.
Only the fiducial volume cut of Section~\ref{subsubsec:fiducial} thins them.

Cosmic rays pass both detectors in coincidence by construction and can escape both cuts above.
A cap on the total BPULSE charge removes them.
It is derived per run as the larger of two values, a \SI{2}{MeV}-equivalent deposit (${\approx}100$~p.e.) or 1.5 times the 99.5th percentile of the coincident charge.
The second value is needed because the $^{60}$Co cascade deposits energy above any cap set for $^{137}$Cs.
Cosmic rays are below $0.1\%$ of the selected sample.

These three cuts define the selected sample of $139{,}527$ events, from which every analysis sample of Section~\ref{subsec:observables} is drawn.
Its waveforms are then processed by the photon counting of Section~\ref{sec:photoncounting}.

\subsection{\textit{In Situ} Single-Photon Characterization}
\label{subsec:spe}

\begin{figure}[t]
    \centering
    \includegraphics[width=\columnwidth]{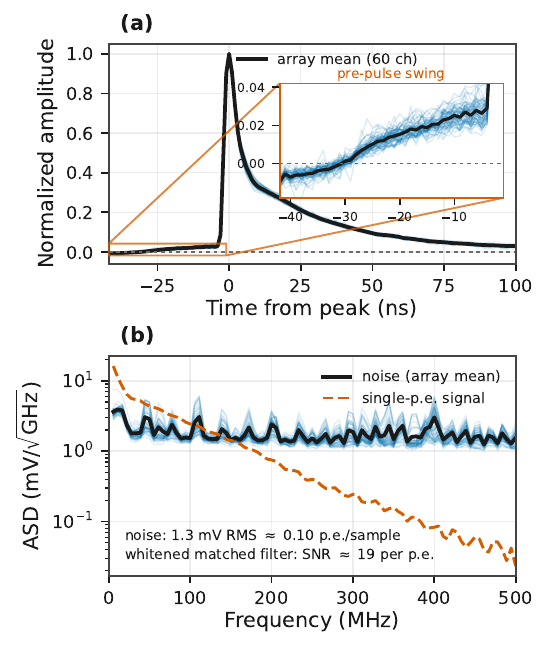}
\caption{\textit{In situ} forward model of the photon-counting fit.
(a) The single photon template $\mathcal{T}(t)$ for every live channel (faint) and the array mean (bold), normalized to unit peak; each channel averages ${\approx}3.5\times10^{3}$ aligned single-p.e.\ snippets, suppressing the single-waveform noise to the sub-percent level seen in the inset.
(b) Amplitude spectral density of the measured noise (per channel and array mean) with the single-p.e.\ signal spectrum at the same normalization. The signal stands a factor ${\approx}5$ above the noise below \SI{50}{\mega\hertz} and crosses the floor near \SI{110}{\mega\hertz}.
Noise dominates each frequency above the crossing, but that band still holds ${\approx}8\%$ of the single-p.e.\ signal-to-noise power and is retained in the fit.}
    \label{fig:spe}
\end{figure}

Counting photons in the selected waveforms requires a per-channel forward model.
Each BPULSE waveform is modeled as a linear superposition of single photon pulses in stationary noise, $w(t) = \sum_{k=1}^{K} a_k\, \mathcal{T}(t - \tau_k) + n$.
The per-photon amplitudes $a_k$ follow a normal distribution about unity, and the noise $n$ has a measured covariance $\boldsymbol{\Sigma}$.
Both the template and the covariance are measured \textit{in situ}. Isolated single-photoelectron pulses, mostly out-of-time light drawn from the waveforms themselves (selection criteria in \ref{app:heuristics}), give $\mathcal{T}(t)$ by alignment and averaging; quiet pre-pulse regions give $\boldsymbol{\Sigma}$ (Fig.~\ref{fig:spe}).
Each amplitude $a_k$ carries a Gaussian prior centered on the single-p.e.\ template amplitude with a relative width of $0.07$, narrower than the $0.10$--$0.19$ per-pulse spread fitted on the same \textit{in situ} sample.
The narrower width is a regularizing choice, and the counts are insensitive to it (\ref{app:rjmcmc-closure}).
The template is averaged over each run and includes the full readout chain, including the ADC calibration and its artifacts, the $4\%$ amplitude spread across the array, the coherent $+4.6\%$ face-to-face offset within it, and the sub-percent gain and temperature drifts of the 18-day campaign.
The per-run, per-channel calibration removes the spread and the offset and averages over the drifts.

From this sample, the mean single-p.e.\ amplitude is \SI{12.0}{\milli\volt} and the mean charge is \SI{186}{\milli\volt\nano\second}, which sets the p.e.\ scale shown in Fig.~\ref{fig:selection}b.
Hits are required to exceed a \SI{6}{\milli\volt} amplitude threshold, half the \SI{12.0}{\milli\volt} single-p.e.\ amplitude.
Integrated charge divided by this mean provides an approximate photoelectron count, which degrades when photons pile up and the time window captures noise and neighboring pulse tails.
With only an approximate scale, the selection variables are quick to compute; the exact counts come from the photon-counting fit.

\subsection{Trans-Dimensional Photon Counting}
\label{sec:photoncounting}

Selected recoils produce per-channel occupancies below ${\approx}10$~p.e.\ (exceeded in $1.2\%$ of channel readings at the brightest operating point), in which the photons overlap heavily within a few tens of nanoseconds.
Integrating charge over a longer window accumulates noise and cannot resolve pileup.
Each photon therefore has to be counted individually.
A trans-dimensional Bayesian deconvolution using reversible-jump Markov chain Monte Carlo (RJMCMC)~\cite{Green:1995} extracts the counts and sub-nanosecond arrival times.
It samples the photon count $K$ and the continuous arrival times $\boldsymbol{\tau}$ together, with amplitudes marginalized analytically against the measured noise covariance.
Each proposed photon adds a dimension, its arrival time, to the fit and must provide enough statistical evidence to justify it.
The method follows the Bayesian single photon waveform analyses developed for photomultipliers~\cite{Xu:2021whl,Wang:2024qxq}. It differs from the published Fast Stochastic Matching Pursuit (FSMP) in its measured colored-noise covariance, flat arrival time priors, and annealed point-estimate output.
It is documented and validated in \ref{app:rjmcmc}, and a fit-independent charge-based estimator corroborates it (Section~\ref{sec:systematics}).
The fit relaxes the selection by processing the full $\Delta t$ range, treating accidental events the same as signal events.

\subsection{Derived Observables and Analysis Samples}
\label{subsec:observables}

The photon counts and arrival times of the per-channel fit yield three observables: an energy from the total count, a vertex from the counts as weights on the fiber positions, and a time response (Section~\ref{subsec:res-timing}).

The fixed \SI{105}{\nano\second} time window runs from \SI{75}{\nano\second} before to \SI{30}{\nano\second} after the 50\% constant-fraction time of the NaI(Tl) pulse, ${\approx}\SI{12}{\nano\second}$ after $t_{\mathrm{NaI}}$.
The total count in this window measures the deposited energy without depending on the occupancy, the number of photons per channel, as a charge integral over the window does (Section~\ref{sec:systematics}).
The prompt signal arrives ${\approx}\SI{42}{\nano\second}$ before the tag, inside this window.
The leading edge of the window, ${\approx}\SI{20}{\nano\second}$ before the prompt, leaves room for the spread of the prompt time relative to the tag.
The trailing edge, ${\approx}\SI{85}{\nano\second}$ after the prompt, spans more than five decay times of the response tail ($\tau \approx \SI{15}{\nano\second}$, Section~\ref{subsec:res-timing}).
A longer window collects more light but also more dark counts.

The center-of-mass vertex weights each fiber position by its photon count (Section~\ref{subsec:res-vertex}).
It estimates the true vertex where the optical confinement is symmetric.
Near a wall the lattice ends on one side, so the estimate is pulled toward the center.
Within the fiducial volume it operates in this symmetric regime, with likelihood-based estimators deferred to future work.

Four analysis samples are used in the results.
\begin{enumerate}
    \item Selected sample: $139{,}527$ events.
    \item Fiducial volume: $3804$ events within the selected sample; restricts the light yield observables to the estimator's well-behaved regime (Section~\ref{subsubsec:fiducial}).
    \item Vertex sample: $126{,}658$ selected events with photons reconstructed on both readout faces and at least four reconstructed photons in total; provides the vertex distribution.
    \item Timing subsample: $36{,}887$ selected events with more than 20 reconstructed photons; provides the arrival time response.
\end{enumerate}

\section{Results}
\label{sec:results}

This section reports the detector response to Compton-tagged \SI{373.4}{keV} electrons at the nominal operating point, $^{137}$Cs at $90^{\circ}$, from the samples in Section~\ref{subsec:observables}.
Table~\ref{tab:kinematics} lists the sample sizes at all four operating points, and Table~\ref{tab:selection} lists the purities.

\subsection{Sensor Response: Dark Rate and Correlated Avalanches}
\label{subsec:res-spe}

The photon-counting fit of Section~\ref{sec:photoncounting} counts the dark and correlated avalanches along with the scintillation photons.
The rates of both are measured from a background run, processed by the same photon-counting fit as the data without the time and charge gates.
Before the rates are computed, the $2.8\%$ of triggers that carry a prompt scintillation flash from the ambient radioactive and cosmic background (at least 5 photons on at least 3 channels within \SI{40}{\nano\second}) are removed.
Dark counts are single avalanches, uncorrelated across the array, at \SI{160}{\kilo\hertz} per channel with a channel-to-channel $\sigma$ of \SI{14}{\kilo\hertz}.
This rate is consistent with the \SIrange{31}{33}{\degreeCelsius} operating point and the temperature dependence of the sensor's dark rate~\cite{NepomukOtte:2016ktf}.

Correlated avalanches are secondary avalanches set off by a primary one, through optical crosstalk to a neighboring cell or through afterpulsing in the same cell.
In the light-yield line of Section~\ref{subsec:res-lightyield}, the dark counts set the intercept and the correlated avalanches scale the slope.
On average they add $2.4\%$ extra avalanches per photoelectron, measured from pairs of avalanches on the same channel of the background run.
Prompt optical crosstalk contributes $P_{\mathrm{CT}} = (0.82 \pm 0.02)\%$, and delayed avalanches within the \SI{105}{\nano\second} time window $P_{\mathrm{delayed}} = (1.55 \pm 0.03)\%$.
Pulse amplitude separates the delayed avalanches into afterpulsing ($0.58\%$) and delayed optical crosstalk ($0.97\%$).
Afterpulses are smaller because the cell has only partly recharged, while delayed crosstalk arrives at full amplitude.
Together they form the correlated avalanche factor applied to the light yield in Section~\ref{subsec:res-lightyield}.

\begin{figure}[t]
    \centering
    \includegraphics[width=\columnwidth]{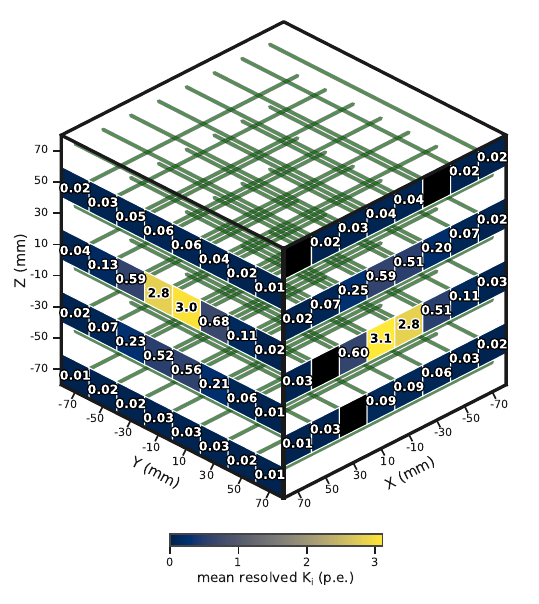}
\caption{Mean p.e.\ per channel over the $3804$ fiducial $^{137}$Cs events, as a corner view of the two instrumented walls seen from outside: the $X$-normal wall facing the tagging spectrometer (left) and the $Y$-normal wall facing the beam (right).
Each cell is one channel at its lattice position; lines trace the WLS fibers and black cells are the four disabled channels.}
    \label{fig:faces}
\end{figure}

\subsection{Event Topology and Vertex Localization}
\label{subsec:res-vertex}

Figure~\ref{fig:faces} shows the number of photons per channel, averaged over the fiducial events, as an image of the collimated beam spot on the fiber lattice.
Light concentrates on the central fibers of both readout faces, consistent with beam gamma rays entering along $+Y$ and depositing energy near the center of the volume.
Each face shows the expected orthogonal projection of one localized energy deposit.

\begin{figure}[t]
    \centering
    \includegraphics[width=0.95\columnwidth]{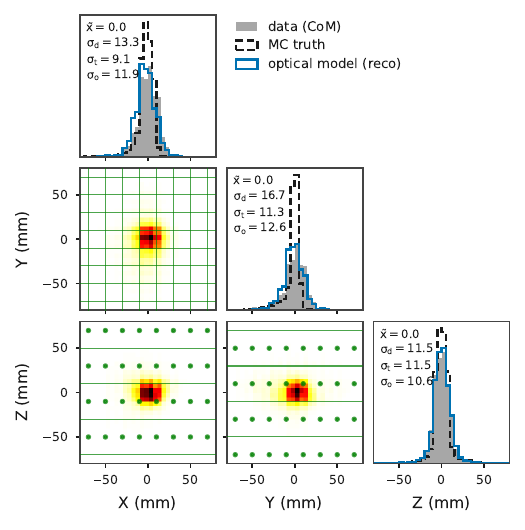}
\caption{Corner plot of the reconstructed center-of-mass vertices of the $^{137}$Cs vertex sample (Section~\ref{subsec:observables}).
Lower triangle: projections onto the detector coordinate planes; green lines and dots mark the fiber lattice at the \SI{20}{\milli\meter} pitch (lines: fibers in the projection plane; dots: perpendicular to it).
Diagonal: marginals overlaid with the energy-weighted centroid of tag-selected transport events (dashed) and with the optical model reconstruction of the same sample (solid; optics consistent with the measured time response).
Annotations list the per-axis median $\tilde{x}$ and the standard deviations of data ($\sigma_d$), simulation truth ($\sigma_t$), and optical model ($\sigma_o$).
The \SI{5}{\milli\meter} bins match the estimator's ${\approx}10$-p.e.-per-face quantization, and finer bins resolve the quantization steps of the estimator, not spatial information.
The simulation truth, with no estimator applied, is shown for scale.
Data are compared directly with the optical model.}
    \label{fig:vertex-corner}
\end{figure}

For every event of the vertex sample, the center-of-mass vertex weights each fiber position by its photon count, coordinate by coordinate over the fibers that resolve it:
\begin{equation}
\hat{r}_i = \frac{\sum_c K_c\, m_{ci}\, r_{ci}}{\sum_c K_c\, m_{ci}}, \qquad m_{ci} = 1 - |n_{ci}|,
\label{eq:com}
\end{equation}
where $K_c$ is the photon count on channel $c$, $\mathbf{r}_c$ the position of its fiber, and $\mathbf{n}_c$ the unit vector along the fiber.
The vertex sample of Section~\ref{subsec:observables} holds $90.8\%$ of the selection.
It applies no lower charge threshold, which would bias the sample toward bright events (Section~\ref{sec:systematics}).
Because the kinematic truth is the beam \emph{axis} and not a point, each axis is summarized by the median $\tilde{x}$ and the standard deviation of the full sample, in the detector frame.

Figure~\ref{fig:vertex-corner} shows the corner plot of these vertices, with the simulation truth, the energy-weighted centroid free of reconstruction resolution, on the diagonal.
The excess of vertices toward the beam entry ($-Y$) confirms the orientation.
The shift of the $X$ distribution toward the tagging spectrometer, ${\approx}\SI{-1.8}{\milli\meter}$ in the mean, is an estimator bias from the four disabled channels, which all sit on the face that resolves $X$.
All three medians are within \SI{0.8}{\milli\meter} of the beam axis; the tagged sample is centered on it.
In the data the widths are $(\sigma_X, \sigma_Y, \sigma_Z) = (13.3, 16.7, 11.5)\,\si{\milli\meter}$, versus $(9.1, 11.3, 11.5)\,\si{\milli\meter}$ in the simulation truth.
The lattice quantization is $\mathrm{pitch}/\sqrt{12} = \SI{5.8}{\milli\meter}$, the standard deviation of a position known only to within one fiber cell.
The measured widths are two to three times this value, so the lattice granularity does not set them.

The optical model re-simulates the tagged transport sample through the geometry with the sensor characterization, the dark rate, optics consistent with the time response, and the light yield reduced to the scale of the data.
Through the same estimator it reproduces the measured $X$ and $Z$ widths to within ${\approx}10\%$ of their values (Fig.~\ref{fig:vertex-corner}), and the interquartile widths, which are insensitive to tails, to model-to-data ratios of $0.93$, $0.97$ and $0.96$.
Over scattering lengths of 1.0--\SI{2.5}{\milli\meter} and absorption lengths of 4--\SI{11}{\meter}, within the grid that covers the time response (Section~\ref{subsec:res-timing}), and both apertures, the reconstructed widths vary by ${\leq}\SI{0.6}{\milli\meter}$ per axis.
The ${\approx}\SI{20}{\milli\meter}$ beam, convolved with the response of the estimator, therefore accounts for the measured widths.
They bound the localization from above without claiming a better resolution.
Separating stochastic confinement from lattice segmentation needs a beam edge sharper than the \SI{20}{\milli\meter} fiber pitch.

The $Y$ marginal also includes the multi-scatter background that passes the energy window, the residual excess of the full sample over the model.
Events off the photopeak sample this background and spread about twice as widely along the beam axis.
The $Y$ width therefore measures the mixture rather than the beam or the resolution.
Mixing the optical-model core with this off-peak distribution ($\sigma_Y \approx \SI{27.3}{\milli\meter}$) at the surviving multi-scatter share ($0.18$ fitted on this sample, $0.25$ expected from Section~\ref{subsec:selection}) gives $\sigma_Y = 16.3$--$17.5~\si{\milli\meter}$.
The observed \SI{16.7}{\milli\meter} falls in this range.

\begin{figure}[t]
    \centering
    \includegraphics[width=\columnwidth]{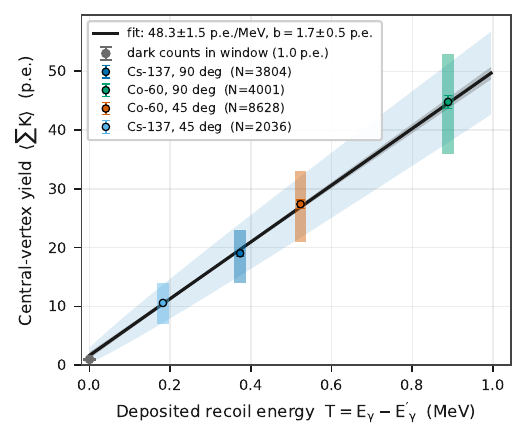}
\caption{Light yield in the fiducial volume against deposited recoil energy at the four operating points.
The line is the fitted straight line and the gray band its $1\sigma$ confidence band.
The marker at $T=0$ is the dark count in the time window (\ref{app:dark-floor}).
Vertical bars on the means combine the statistical uncertainty with the $\pm 2.5\%$ run-to-run systematic (Table~\ref{tab:systematics}); the $\pm 4$~keV systematic on the mean energy is smaller than the markers.
The light band is the Poisson counting width about the fit, and the colored band at each point is the measured 16--84\% width of the events.}
    \label{fig:lightyield}
\end{figure}

\subsection{Light Yield and Linearity}
\label{subsec:res-lightyield}

The central light yield, the mean total reconstructed photon count of the $3804$ fiducial events after accidental subtraction and with no charge threshold, is $19.04 \pm 0.07$~p.e.\ at \SI{373.4}{keV}.
Throughout, p.e.\ refers to counted avalanches.
The photoelectron yield, corrected for correlated avalanches, is given only where stated.

Including the $^{60}$Co points and the \SI{182}{keV} $^{137}$Cs point, the response is consistent with a straight line at $48.3 \pm 1.5$~p.e.\ per MeV (Fig.~\ref{fig:lightyield}).
We model the response with a straight line fitted with uncertainties on both axes, assigning each yield its statistical uncertainty combined with the $\pm 2.5\%$ run-to-run systematic of Table~\ref{tab:systematics} and each deposited energy the ${\approx}\SI{4}{keV}$ intrinsic energy width.
The two unresolved $^{60}$Co lines broaden the count distributions but not the mean deposited energies.
Under a linear response, the mean yield then tracks the mean energy deposit closely.
Once the mean deposited energies are fixed, only the tag energy scale and acceptance contribute to the uncertainties of these energies.
The fit gives $\chi^2 = 2.26$ for two degrees of freedom ($\chi^2/\mathrm{dof} = 1.13$, $p = 0.32$).
The quoted uncertainties include the scaling by $\sqrt{\chi^2/\mathrm{dof}}$.
The fitted intercept, $1.7 \pm 0.5$~p.e., is consistent with the dark counts in the time window plus the residual of the selection bias rather than with a response offset.
The dark counts alone contribute ${\approx}1.0$~p.e.\ (marked on the figure and derived in \ref{app:dark-floor}).
Unlike the intercept, the slope in counted avalanches includes all correlated avalanches.
After division by the correlated avalanche factor ($1 + P_{\mathrm{CT}} + P_{\mathrm{delayed}} = 1.024$), the photoelectron yield of the single-ended readout, in its current form, is $47.2 \pm 1.5$~p.e.\ per MeV\@.
Section~\ref{sec:discussion} estimates possible improvements to the light yield.

The coincident cascade gamma of $^{60}$Co adds a second in-time energy deposit that would bias single-vertex imaging and first photon timing.
The \SI{182}{keV} point gives too little light for either analysis; the $45^{\circ}$ beam geometry illuminates a different volume.
On the other runs, the vertex and timing analyses reproduce the response at the nominal point where the beam geometry is shared, with vertex widths within \SI{0.8}{\milli\meter} in $X$ and $Y$ and timing precision within $3.5\%$ of the nominal-point values.
The residual broadening of the $Z$ width comes from the cascade gamma.

The four points deviate from the fitted line by $0.5$--$3\%$.
A constant offset changes the intercept but not the slope.
Quenching is tested by the per-MeV yield ratios between the $^{137}$Cs and $^{60}$Co points at each angle (\ref{app:self-selection}), with the background inside the coincidence window subtracted.
At their $2$--$3\%$ precision the ratios are consistent with a linear response.
They are also consistent with the Birks coefficients fitted for LAB scintillators, $k_B \approx 0.012$--$0.015~\mathrm{g\,cm^{-2}\,MeV^{-1}}$~\cite{JUNO:2024fdc,DayaBay:2019fje}, which imply deficits of $2$--$4\%$ at these energies.
The same ratios exclude $k_B > 0.03~\mathrm{g\,cm^{-2}\,MeV^{-1}}$ at $95\%$ confidence.
The four-point fit alone constrains quenching only weakly, because over 182--\SI{963}{keV} a Birks response~\cite{Birks:1951} resembles a small change of slope and intercept.
A second nonlinearity, the Cherenkov threshold for electrons in the LAB matrix, falls at the \SI{182}{keV} point and below the three higher points.
Cherenkov light is a small fraction of the total light collected, so its onset is also below the sensitivity of this measurement.

\subsection{Energy Resolution}
\label{subsec:res-resolution}

The resolution at the central yield is $\sigma/E = (23.4 \pm 0.4)\%$, consistent with the Poisson counting expectation $1/\sqrt{N_{\mathrm{pe}}} = 22.9\%$.
This width is the Poisson spread of the photoelectron count within each event; the statistical uncertainty of the event sample is the $\pm 0.4\%$.
It is defined as the normalized median absolute deviation over the mean ($1.4826\times\mathrm{MAD}$), so that the multi-scatter and pileup tails do not enter.

The shape of the count distributions tests whether the resolution is Poisson-limited.
The fiducial spectrum of each operating point is compared to a Poisson mixture with a structure fixed by the calibration setup.
The collimator and slit geometry hold the recoil width of each accepted line at the ${\approx}\SI{4}{keV}$ of Section~\ref{subsec:bench}.
Each line contributes a pure Poisson term at its mean, while the NaI(Tl) resolution affects only the weights of the lines in the energy window.
Once the correlated avalanche branching and the dark counts in the time window are included, the light yield scale is the single fitted parameter.
If the selection accepts events with a probability that varies exponentially with $K$, a Poisson distribution remains Poisson with a shifted mean.
The selection bias then changes only the fitted scale.
Any other dependence on $K$ would instead appear as a shape failure that the $\chi^2$ of the mixture comparison bounds.
In 2-p.e.\ bins over the central $98\%$ of the mixture, the reduced Pearson $\chi^2/\nu$ is $1.21$, $1.14$, $1.30$, and $1.32$ with increasing recoil energy ($^{137}$Cs at $45^{\circ}$ and $90^{\circ}$, then $^{60}$Co at $45^{\circ}$ and $90^{\circ}$), each within one standard deviation $\sqrt{2/\nu}$ of unity ($p = 0.30$, $0.33$, $0.21$, $0.17$).
The counting is Poisson-limited at all four operating points.
The odd--even jitter of the single-p.e.\ bins comes from the integer sums per channel, repeats in all four runs, and cancels on the 2-p.e.\ bins.
Figure~\ref{fig:lightyield} compares the observed 16--84\% width of each distribution with the counting band.
Relative to the $\pm1\sigma$ expectation of the mixture, the intervals agree to better than $1$~p.e.\ at every operating point.
In particular, the $^{60}$Co widths exceed the pure counting band by the predicted line mixing term.
The remaining excess sits in the tails, all of them known populations.
Outside the core, these populations add up to less than $3\%$.
The excess of high-count events over the mixture comes from coincidence summing with the cascade gamma at the two $^{60}$Co points.
On the low side, partial in-time energy deposits dominate at the three brighter points, while at \SI{182}{keV} the low-$K$ edge of the selection bias dominates.
The accidental background, measured in the off-time sideband, is below $0.2\%$ of the fiducial sample at every operating point.
Within its single-scatter core, the brightest point passes the same $\chi^2$ test as the other three ($p = 0.17$).

\begin{figure}[t]
    \centering
    \includegraphics[width=\columnwidth]{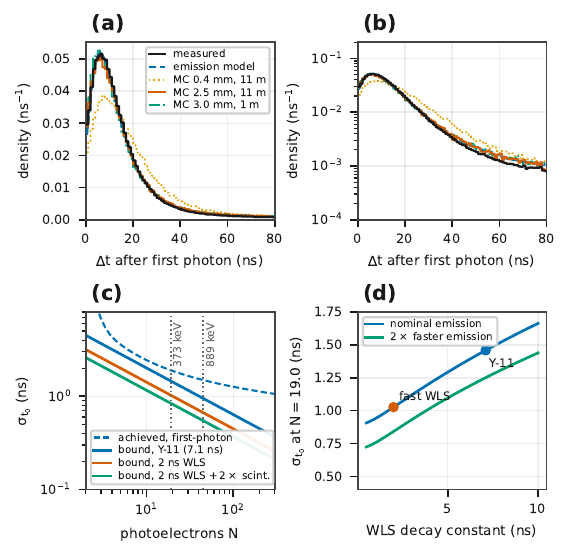}
\caption{Timing response and its design reach.
(a), (b) First-photon-referenced time response of the timing sample ($36{,}887$ events), overlaid with the analytic emission model and with transport-only Monte Carlo kernels (delta emission reconvolved with the same emission model, no instrumental term).
Three kernels span the response: (\SI{0.4}{\milli\meter} scattering, \SI{11}{\meter} absorption) undershoots the peak, (\SI{2.5}{\milli\meter}, \SI{11}{\meter}) matches, and (\SI{3.0}{\milli\meter}, \SI{1}{\meter}) overshoots.
(c) Start-time resolution as a function of collected photoelectrons: dashed, achieved by the first-photon time of this analysis after its bias calibration; solid, the Cram\'er--Rao bound for the emission chain (blue) and two cumulatively faster variants.
Vertical lines mark the measured mean counts of the two tagged energies.
(d) The timing lower bound (Cram\'er--Rao) at $N = 19.0$ as a function of the WLS re-emission constant.}
    \label{fig:time-response}
    \label{fig:crlb}
\end{figure}

\subsection{Timing Response}
\label{subsec:res-timing}

The earliest reconstructed photon over all channels, after bias calibration, sets the event time, called the first-photon time.
Figure~\ref{fig:time-response} shows the photon arrival times of the tagged sample relative to it, the time response.
During the construction of the kernel, a per-event start time fit with a Poisson likelihood on the dark-count rate removes the dark counts that precede the fitted start; later ones remain and are modeled.
The start-time fit is stable above 20 reconstructed photons, which sets the threshold of the timing subsample of Section~\ref{subsec:observables}.
On this sample the response peaks near \SI{7}{\nano\second} and decays with $\tau \approx \SI{11}{\nano\second}$ over 8--\SI{25}{\nano\second} and with $\tau \approx \SI{15}{\nano\second}$ over 25--\SI{60}{\nano\second}.
An analytic model combines a four-component LAB+PPO+bis-MSB emission profile~\cite{Beretta:2025jinst} with the \SI{7.10}{\nano\second} Y-11 re-emission constant~\cite{Kodama:2023suc} and a \SI{2.0}{\nano\second} Gaussian.
The Gaussian is an empirical term that fixes the model to the measured distribution, not a measured quantity.
The model reproduces the observed curve through the first-photon time, so the decay is dominated by the scintillator emission and the Y-11 re-emission.
With the transport term set to zero, the Gaussian accounts for all the per-photon dispersion left unexplained by the emission model, whether from the readout or from optical transport.
Independent measurements constrain the emission constants of the model, which set the decay.

A second simulation of the tagged sample through the optical model with instantaneous emission produces the transport kernels.
They are subsampled to the counts of the data.
The distribution of the first-photon time, an order statistic (the earliest of the photon times), depends on the number of photons drawn from the kernel.
With scattering lengths ranging over 0.4--\SI{3.0}{\milli\meter} and absorption lengths over 1--\SI{11}{\meter}, the kernels cover the observed response (Fig.~\ref{fig:time-response}a,b).
Over this grid of scattering and absorption lengths, with the true values not yet measured, the peak height of the response varies by $25\%$.
The capture length of the fiber, not varied here, would change the kernel in the same way as the scintillator transport, so the two cannot be separated.

The readout does not explain the \SI{2.0}{\nano\second} dispersion.
Among the four fibers closest to the beam spot, the offsets within one face are at most $0.15$~ns, and the coherent ${\approx}\SI{0.8}{\nano\second}$ offset between the two digitizer halves of Section~\ref{subsec:lowlevel}, one per face, is a split of $\pm 0.4$~ns about the common mean.
The $D_{2d}$ symmetry of the lattice makes their transport paths equivalent once the beam spot offset is regressed out.
Without the regression, the residual beam spot geometry increases the spread of the same channels to a conservative ${\lesssim}\SI{0.5}{\nano\second}$ RMS\@.
Sample composition does not explain it either.
The timing sample includes no single-scatter requirement, yet a fit of the response to its fiducial subset ($1398$ events, single-scatter purity ${\approx}90\%$) returns $\sigma = 2.08 \pm 0.05$~ns compared to $1.98 \pm 0.01$~ns for the full sample, with the peak and decay shapes unchanged.
The dispersion therefore reflects optical transport that the emission model leaves out, not the instrument.
Like the overall light collection scale of the simulation, it awaits a measurement of the optical constants.
Wherever the emission model is used, the Gaussian stands in for it.

The validated emission model also guides timing design.
Consider $N$ photons drawn independently from the single-scatter response $f(t)$, where $N = \sum_i K_i$ is the total photon count of the event.
The Cram\'er--Rao bound on their common start time is $\sigma_{t_0} \geq [\,N \int (f'^2/f)\,dt\,]^{-1/2}$~\cite{Seifert:2012pmb}.
On the model that reproduces Fig.~\ref{fig:time-response}, this timing bound is $\sigma_{t_0} \approx \SI{1.5}{\nano\second}$ for the system at the nominal operating point ($N = 19.0$, Fig.~\ref{fig:crlb}c).
The first-photon time gives \SI{1.90}{\nano\second}, about $30\%$ above the bound.
Pseudo-experiments that draw $N = 19$ photons from this response and fit the full likelihood over all arrival times realize it to within $5\%$ once the constant \SI{0.3}{\nano\second} shift of their fitted start time, a bias of the estimator, is calibrated out.
Because every photon arrival is resolved, the full likelihood can be evaluated.
The first-photon time uses only the earliest photon, whose spread at high counts is set by the tail of the \SI{2.0}{\nano\second} Gaussian and shrinks only slowly with $N$.
The full likelihood uses every photon, so its advantage over the first-photon time grows at higher counts.
In detectors with parallel fibers, such as AntiMatter-OTech, the coordinate along the fibers comes from these arrival times, so this response and its bound set the resolution of that coordinate; here the orthogonal lattice measures it directly.
For alternative materials, the bound quantifies the timing upgrades of the design discussion; scaling as $1/\sqrt{N}$, it improves with the square root of the light yield.

\section{Systematic Uncertainties}
\label{sec:systematics}

Once the selection bias of the fiducial cut is corrected, every term in the budget of Table~\ref{tab:systematics} is at or below the few-percent level.
Each entry below states the bound and how it was measured; derivations longer than a few sentences are collected in \ref{app:selection-derivations}.

The NaI(Tl) energy calibration is validated against the Compton kinematics of both sources.
At $^{137}$Cs $90^{\circ}$ the tagged photopeak reconstructs at \SI{288.0}{keV} against the \SI{288.3}{keV} expectation.
The three other operating points agree to 1.0--1.3\% with expectations that fold in their acceptance, the largest difference being 736.5 versus \SI{727.4}{keV} at the $^{60}$Co $45^{\circ}$ doublet.
The same 288--\SI{737}{keV} agreement limits the non-proportionality of the NaI(Tl) crystal to 1--2\% below ${\approx}\SI{400}{keV}$.
The resulting uncertainty on $E_e$ is at most $2.6\%$, reached at the smallest recoil energy.

The photon-counting fidelity of Table~\ref{tab:systematics}, how often the fit returns the true number of photons, rests on the synthetic closure of \ref{app:rjmcmc-closure} and on a cross-check independent of the fit.
In the cross-check, the heuristic pulse charges are integrated over the \SI{105}{\nano\second} time window and divided by the single-p.e.\ charge of each channel.
The charge estimate follows the photon count event by event with a correlation of $0.98$.
Its ratio to the count is 1.12--1.15 at the operating points and rises with occupancy, as overlapping pulses add tail charge to the integral; at the lowest occupancies, where pulses are resolved in time, the two agree.
The counted response does not share this dependence, because each photon contributes one reconstructed pulse rather than integrated charge.

\begin{table}[t]
\centering
\caption{Systematic-uncertainty budget on the light yield and energy-response results, grouped by the affected quantity; the notes give the supporting arithmetic.}
\label{tab:systematics}
\footnotesize
\setlength{\tabcolsep}{3pt}
\begin{tabular}{@{}>{\raggedright\arraybackslash}p{0.55\columnwidth}>{\raggedright\arraybackslash}p{0.40\columnwidth}@{}}
\toprule
Source & Magnitude \\
\midrule
\multicolumn{2}{@{}l}{\itshape Yield scale} \\
Energy scale (intrinsic energy width) & ${\approx}1\%^{\,a}$ \\
NaI(Tl) energy calibration (propagated as $E_{\gamma}'/E_e$) & ${\leq}2.6\%$ on $E_e$  \\
Subdominant terms (accidental subtraction, coincidence window, gain drift, multi-scatter quenching) & each $<1\%^{\,b}$ \\
Correlated-avalanche correction ($1{+}P_{\mathrm{CT}}{+}P_{\mathrm{delayed}}$, applied) & $2.4\%$ ($0.8$ prompt $+$ $1.6$ delayed: afterpulse $+$ delayed crosstalk) \\
Photon-counting fidelity & ${\leq}2\%$ miscount at 12~p.e.\ (\ref{app:rjmcmc-closure}) \\
Multi-scatter background under the tag peak & ${\lesssim}3\%$ on the mean$^{\,c}$ \\
Selection bias of the fiducial cut & ${\leq}2.3\%$ on the slope; $+14$--$16\%$ on the lowest-point mean before correction at a common light level  \\
\midrule
\multicolumn{2}{@{}l}{\itshape Fit intercept} \\
In-window dark counts & $0.99 \pm 0.05$~p.e./event, common to all points  \\
\midrule
\multicolumn{2}{@{}l}{\itshape Four-point linearity} \\
Operating-point comparability & $\pm 2$--$3\%$; $\pm 2.5\%$ adopted$^{\,d}$ \\
$^{60}$Co coincidence summing & $+1.3\%$ one-sided on the $^{60}$Co means  \\
\bottomrule
\end{tabular}\\[2pt]
{\footnotesize\raggedright
$^{a}$~The ${\approx}\SI{4}{keV}$ intrinsic energy width of Section~\ref{subsec:bench}.\par\smallskip
$^{b}$~Substituting the far-$\Delta t$ plateau ($\Delta t < -\SI{80}{\nano\second}$) for the near sideband changes the central mean by $<0.1\%$, and widening the coincidence window to $\pm\SI{14}{\nano\second}$ changes it by $<0.4\%$. Gain drift after per-channel calibration is measured at $<1\%$. The multi-scatter light deficit from quenching follows from $10\%$ admixture $\times$ ${\approx}10\%$ secondary-energy fraction $\times$ a few- to ten-percent difference in quenching, or ${\lesssim}0.1\%$, at $(90.4 \pm 0.8)\%$ single-scatter purity.\par\smallskip
$^{c}$~Measured by passing tag amplitudes outside the photopeak through the production fit, of which the fiducial cut passes $1.2\%$ ($0.7\%$) of the low (high) band.\par\smallskip
$^{d}$~NoWaSH transport productions at the two angles: multi-scatter fraction before the fiducial cut $15.0\%$ at $45^{\circ}$ and $24.8\%$ at $90^{\circ}$; single-scatter purity after the cut $(92.9 \pm 0.5)\%$ and $(90.4 \pm 0.8)\%$.\par}
\end{table}

No charge threshold is applied at any operating point.
In the off-time sideband, $82\%$ of events are at or below 2~p.e.
These low-count events are mostly accidental coincidences, and the off-time subtraction already removes them, so a threshold is not needed to suppress them.
In a threshold scan, requiring at least ${\approx}8$~p.e.\ per event truncates the Poisson fluctuations of the signal and biases the \SI{182}{keV} mean by $+16\%$.

Because the fiducial cut acts on a vertex reconstructed from few photons, a dim event passes it more often when its photon count fluctuates upward.
This selection bias raises the accepted mean by $+14$--$16\%$ at the lowest operating point (\ref{app:self-selection}).
Common to all four points, it cancels in the slope at the ${\leq}2.3\%$ level of Table~\ref{tab:systematics}.

The multi-scatter background surviving the fiducial cut could alter the mean yield in two ways.
It makes up $(8 \pm 3)\%$ of the fiducial events, as measured from the pedestal under the peak of the tag spectrum.
Its light barely moves the mean.
Fiducial events in the tag bands on either side of the photopeak, which sample this background, average $21.2$ and $13.7$~p.e., within $5.3$~p.e.\ of the $19.0$~p.e.\ signal.
Even at the upper end of its share, the background therefore shifts the mean by ${\lesssim}3\%$.
Energy conservation makes its energy deposit degenerate with the signal, leaving only the light deficit from its quenching.

The dark counts in the time window are measured with the same reconstruction used for the data (Section~\ref{sec:processing}).
Their level is $0.99$~p.e.\ per event, from the dark rate of Section~\ref{subsec:res-spe} (\ref{app:dark-floor}), and is the same at every operating point.
The photon density before the prompt signal in the tagged $^{137}$Cs events confirms it to $5\%$.
Being common to all points, it shifts the intercept of the response fit rather than its slope.
The $^{60}$Co runs also carry a few tenths of a p.e.\ of light from a second gamma ray in the time window, carried in Table~\ref{tab:systematics} as $^{60}$Co coincidence summing.

The single-scatter purity after the fiducial cut differs between the two angles by $2.5$ percentage points with the \SI{3}{\milli\meter} slit and $3.5$ with the \SI{4}{\milli\meter} slit (Section~\ref{subsec:expected-signal}; Table~\ref{tab:systematics}, note d).
We adopt $\pm 2.5\%$ as the comparability band.
Direct measurement of the coincident cascade gamma fraction puts its effect at $+1.3\%$ on the $^{60}$Co means, equivalent to a $-1.1\%$ shift of the four-point slope if corrected.
Table~\ref{tab:systematics} lists it as a one-sided uncertainty rather than a correction to the $^{60}$Co points.

The multi-scatter admixture of Table~\ref{tab:systematics} accounts for most of the tails outside the fitted core.
The residual is the coincidence-summing tail and the low-count edge of the selection, both identified in the mixture comparison.

\section{Discussion: Implications for Detector Design}
\label{sec:discussion}

\begin{figure}[htbp]
    \centering
    \includegraphics[width=\columnwidth]{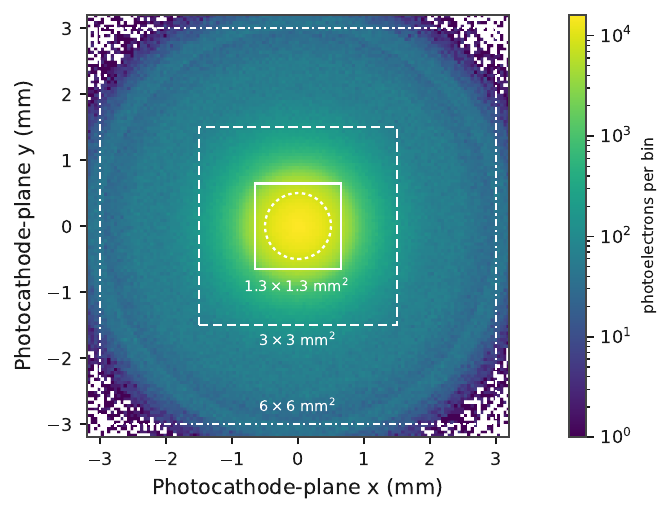}
\caption{Simulated photon arrival distribution at the SiPM plane for a single fiber--sensor coupling (Geant4 model of the grease-filled gap and epoxy window; logarithmic color scale).
Photons exiting the \SI{1}{\milli\meter} fiber (dotted circle) disperse through the grease-filled gap and module window beyond the $1.3 \times 1.3~\mathrm{mm}^2$ baseline active area (solid square) and the $3 \times 3$ and $6 \times 6~\mathrm{mm}^2$ enlargements (dashed); the capture ratios are quoted in the text.}
    \label{fig:sensor-coupling}
\end{figure}

Because the resolution in this testbed is set by the photoelectron count (Section~\ref{subsec:res-resolution}), the energy resolution of a full-scale detector would be set by its light collection if that holds at scale.
Our ${\approx}48$~p.e.\ per MeV is the starting point for projections toward the higher light yields that planned LiquidO detectors assume, for the Compton-vertex imaging of COCOA~\cite{LiquidO:2025hbe} and for NuDoubt++~\cite{NuDoubt:2024jax}.
CLOUD uses the same kind of WLS-fiber and SiPM readout, with both fiber ends instrumented~\cite{NavasNicolas:2024ixn}.
Two readout changes could raise the light yield: reading out both fiber ends and enlarging the sensor.
Dual-ended readout, assumed in the simulation geometry, would recover the backward-going half of the confined light that the far ends under the black cap currently absorb.
At this fiber length, short compared to the attenuation length, the recovery is a full factor of two.
A mirrored far end, standard in single-ended scintillator trackers, would recover most of it without additional channels.
Dual-ended readout would also make the position measurements from the lattice and from the difference in arrival time at the two fiber ends independent, so the better position measurement could validate a timing algorithm event by event.
A larger sensor helps because the present one intercepts only ${\approx}80\%$ of the light arriving at the sensor plane.
Photons exiting the fiber spread through the grease-filled coupling gap and the module window well beyond the $1.3 \times 1.3~\mathrm{mm}^2$ active area at the \SI{0.5}{\milli\meter} standoff of Section~\ref{subsec:daq}.
In a Geant4 model of the coupling (Fig.~\ref{fig:sensor-coupling}), enlarging the active area to $3 \times 3~\mathrm{mm}^2$ ($6 \times 6~\mathrm{mm}^2$) gives $1.16\times$ ($1.21\times$) the baseline detected light yield.
Together the dual-ended readout and the $6 \times 6$ sensor project the response to 108--124~p.e.\ per MeV; the range covers a \SIrange{0.05}{0.30}{\milli\meter} assembly tolerance of the coupling gap.

The Cube prototype~\cite{LiquidO:2025rhe}, which has both upgrades with both ends of all 64 fibers (128 channels) read out by $3 \times 3~\mathrm{mm}^2$ air-coupled sensors, gives a direct cross-check of this projection.
It reports ${\approx}170$~p.e.\ per MeV (${\pm}40\%$, dominated by its time-over-threshold charge conversion) for cosmic-muon tracks in a 15\,wt\% wax NoWaSH volume.
The projection lies within that band, so the factor ${\approx}3.5$ between the two responses is attributed to the fiber ends and sensor coupling rather than to the target medium.
Differences remain in the fiber type and pitch (BCF-91A at \SI{3.2}{\milli\meter} versus Y-11 at \SI{20}{\milli\meter}), in track-like rather than point-like energy deposits, and in the air-gap sensor coupling of the Cube.
In this model, the air gap of the Cube collects ${\approx}30\%$ less light than the grease coupling of BPULSE\@.
The comparison is consistent, but the ${\pm}40\%$ uncertainty of the Cube measurement limits its precision.

Because the dark count rate grows with the junction area, the per-channel rate of a $6 \times 6~\mathrm{mm}^2$ sensor becomes ${\approx}\SI{3.4}{\mega\hertz}$, raising the dark counts in the time window from ${\approx}1$ to ${\approx}21$~p.e.
At the \SIrange{31}{33}{\degreeCelsius} sensor temperature of this campaign, this background cancels the full statistical gain, degrading the energy resolution at \SI{373.4}{keV} from the ${\approx}21\%$ that the light gain of the enlarged sensor alone would deliver to ${\approx}29\%$.
Cooling the sensors to about \SI{10}{\degreeCelsius} recovers it, because the dark count rate roughly halves for every \SIrange{5}{7}{\degreeCelsius} in this sensor class~\cite{NepomukOtte:2016ktf}.
A larger sensor without cooling gains no resolution.
The intermediate $3 \times 3~\mathrm{mm}^2$ step adds only ${\approx}5$~p.e.\ of dark counts.

In the same model the grease coupling collects $45\%$ more light than an otherwise-identical air gap, so the coupling medium offers no further gain.
The remaining loss is geometric, set by the ${\approx}\SI{0.5}{\milli\meter}$ of grease and epoxy between the fiber and the silicon.
A thinner gap would narrow the arrival pattern with no dark-rate penalty.
The epoxy window, however, is part of the packaged SiPM and cannot easily be thinned without risk to the sensor.
At this standoff a $6 \times 6~\mathrm{mm}^2$ sensor collects ${\approx}96\%$ of the light arriving anywhere on its face.
A larger sensor already captures nearly all of this light, so a thinner gap adds little to it.
A higher dye loading, shortening the capture length, would plausibly improve both collection and the arrival time spread, though we leave that path unquantified.

The vertex and timing results carry two further implications for a larger opaque tracker.
At the \SI{20}{\milli\meter} fiber pitch, the measured vertex widths of 11.5--\SI{16.7}{\milli\meter} per axis are set by the beam spot rather than by the pitch (Section~\ref{subsec:res-vertex}).
For timing, the Cram\'er--Rao bound on the single-scatter start time of the system is ${\approx}\SI{1.5}{\nano\second}$ for the nominal $19.0$~p.e.
This bound comes from the Y-11 re-emission time constant of \SI{7.10}{\nano\second}~\cite{Kodama:2023suc} and the \SI{2.0}{\nano\second} transport dispersion, not from the photoelectron count alone.
With readout offsets ${\lesssim}\SI{0.5}{\nano\second}$, a per-channel delay calibration would gain little against the limit set by the wavelength shifter.
Fast wavelength shifters with ${\approx}\SI{2}{\nano\second}$ decay and emission nearly matched to the Y-11 band are now available (Kuraray YS-4, $2.15 \pm 0.03$~ns)~\cite{Kodama:2023suc}.
One such substitution lowers the bound to \SI{1.03}{\nano\second} at the same $19.0$~p.e.
With a faster scintillator formulation as well, the bound would approach \SI{0.84}{\nano\second} at that light level (Fig.~\ref{fig:crlb}d).
A higher light yield lowers each of these bounds further, as $1/\sqrt{N}$.

The beam spot size limits the vertex comparison, and thus an important upgrade to the calibration setup is a narrower collimator bore and aperture slit.
For sizes below the current aperture, the machining precision of the slit edges and their alignment would set the width of the beam spot.
A narrower beam spot separates the beam profile from the detector resolution and could support future studies constraining the optical properties of the target medium.

A narrower beam spot lowers the event rate.
A second tagging arm would offset that loss by recording two angles in a single acquisition, since the operating points differ only in the angle of the arm (the spectrometer, aperture, and shielding assembly).
The geometry can accommodate several arms at once.
LaBr$_3$:Ce is an off-the-shelf replacement for the spectrometer.
Its ${\approx}3\%$ energy resolution at \SI{662}{keV}~\cite{van_Loef_2001} (versus ${\approx}7\%$ for NaI(Tl)~\cite{Moszynski:2002kr}) narrows the energy window ${\approx}2.3\times$ at fixed photopeak efficiency.
A narrower window accepts less of the multi-scatter continuum under the tag peak.
At ${\geq}5\%$ Ce doping its principal decay time of about \SI{17}{\nano\second} and its sub-nanosecond rise~\cite{Glodo_2005} sharpen the coincidence.
Because that continuum is nearly flat across the peak, the accepted background shrinks by about the same factor.
The tag purity then rises from $0.72$ to ${\approx}0.86$ at unchanged efficiency.
The single-scatter fraction, fixed by the target physics, is unaffected.

\section{Conclusions}

BPULSE was built to close a specific gap: the response of an opaque, stochastically confining scintillator to isolated sub-MeV electrons at the center of the target had never been measured against an independent kinematic ground truth.
Compton tagging of $45^{\circ}$ and $90^{\circ}$ scatters of collimated gamma-ray beams from the $^{137}$Cs and $^{60}$Co sources isolates recoils of known energy, and the light is counted channel by channel with a trans-dimensional fit based on an \textit{in situ} single photon model.
The detector collects $19.04 \pm 0.07$ photoelectrons for \SI{373.4}{keV} recoils.
Across four operating points the fitted response is $48.3 \pm 1.5$~p.e.\ per MeV, consistent with a linear response whose intercept is accounted for by the dark counts and the selection residue.
With both fiber ends read out and larger sensors, simulated gain factors project this target to 108--124~p.e.\ per MeV, consistent with the light yield of the dual-ended Cube prototype.
At the central yield the resolution, $(23.4 \pm 0.4)\%$, is consistent with the Poisson expectation of $22.9\%$, and at all four operating points the count distributions are Poisson mixtures.
The center-of-mass vertex places the tagged interactions on the beam axis, and the spread of the reconstructed vertices matches the width of the beam spot.
This width bounds the response set by light confinement from above, as the optical model benchmark confirms.
The arrival time response decays with the emission constants.
Its Cram\'er--Rao bound on the single-scatter start time, ${\approx}\SI{1.5}{\nano\second}$ at this light level, is set by the re-emission of the wavelength shifter rather than by the photoelectron count alone.
The peak of the response is directly sensitive to the millimeter-scale scattering length and to the absorption length, neither yet measured.

Three directions excluded by design from this first campaign are now within reach.
The opaque water-based liquid scintillator (oWbLS)~\cite{LiquidO:2024piw,Che:2026wab}, an opaque formulation with scattering lengths of a few millimeters, is the subject of an ongoing campaign.
Multi-scatter topologies, rejected by the single-scatter selection of this campaign, are the direct test of the tracking capability of the technology.
The recorded waveforms already contain the time-separated pairs of energy deposits within single events that a dedicated analysis, now in preparation, will quantify.
Directionality is accessible through this tagged geometry at additional scattering angles.
An extended angular scan will measure the response across the full sub-MeV range, with the $^{60}$Co operating points at its high-energy end.

\section*{Acknowledgments}

We thank the Department of Physics and Nuclear Engineering at the United States Military Academy, West Point, for the use of their gamma sources, shielding, and laboratory space.
We thank Dat Man Ho, University of Michigan, for machining the acrylic detector vessels.
We thank the LiquidO group at the University of Sussex for their hospitality during extended visits and for many discussions.
This work was supported by the Department of Energy, Nuclear Energy University Program Fellowship, United States; the Department of Defense, Defense Threat Reduction Agency (HDTRA1-20-2-0002); the Department of Energy, National Nuclear Security Administration, Consortium for Monitoring, Technology, and Verification (DE-NA0003920); and the Applied Research Lab at Penn State.
Brookhaven National Laboratory is supported by the U.S. Department of Energy under contract DE-AC02-98CH10886.
We acknowledge the U.S. National Science Foundation Major Research Instrumentation Program (PHY-2018280), Deep Learning for Statistics, Astrophysics, Geoscience, Engineering, Meteorology and Atmospheric Science, Physical Sciences and Psychology (DL-SAGEMAPP) at the Institute for Computational and Data Sciences (ICDS) at the Pennsylvania State University.
The LiquidO Collaboration gratefully acknowledges support in the UK by STFC (grants ST/S000798/1, ST/W005700/1, ST/V001361/1, ST/W000512/1, ST/X006026/1, ST/X004961/1) and the Royal Society (grants SRF/R1/231053, RSWVF/R3/223011), and in Germany by the Cluster of Excellence ``Precision Physics, Fundamental Interactions, and Structure of Matter'' (PRISMA$^{+}$ EXC 2118/1 and PRISMA$^{++}$ EXC 2118/2) funded by the German Research Foundation (DFG) within the German Excellence Strategy (Project ID 390831469).
We gratefully acknowledge the support of the CNPq/CAPES in Brazil, the McDonald Institute providing FVRF support in Canada, the Charles University in the Czech Republic, the CNRS/IN2P3 in France, the INFN in Italy, the Funda\c{c}\~{a}o para a Ci\^{e}ncia e a Tecnologia (FCT) in Portugal, the CIEMAT in Spain, and the University of California at Irvine for their provision of resources.
The research leading to these results has received funding from the Spanish Ministry of Science and Innovation PID2022-136297NB-I00/AEI/10.13039/501100011033/FEDER, UE\@. IFAE is partially funded by the CERCA program of the Generalitat de Catalunya.

\appendix

\section{The Trans-Dimensional Waveform Fit}
\label{app:rjmcmc}

The photon-counting fit is summarized in Section~\ref{sec:photoncounting}.
This appendix describes the signal model, search, and implementation choices that evaluate the energy without approximation and keep the search fast enough to fit every waveform in the selection, ${\approx}\SI{6}{\milli\second}$ per waveform (\ref{app:profile}).
It begins with the heuristic pulse finding used to prepare the waveforms and select the single photon calibration sample.
These implementation choices determine the reproducibility of the results and the practicality of the ungated dark-noise analysis of Section~\ref{subsec:res-spe}.

\subsection{Heuristic Pulse Finding and Single Photon Candidate Selection}
\label{app:heuristics}

For pulse finding, a copy of each waveform is passed through a \SI{150}{\mega\hertz} low-pass filter.
Its baseline is estimated by a median absolute deviation procedure that iteratively excludes the samples of candidate pulses.
A threshold scan on the filtered trace identifies candidate pulses.
The local minima bounding each pulse mark its integration bounds.
A $50\%$ constant-fraction discriminator interpolates each sub-sample arrival time.
Single-photon candidates for the template of Section~\ref{subsec:spe} are selected from the two-dimensional distribution of pulse charge versus amplitude.
A candidate is accepted inside a contour drawn around the single-photoelectron population of that distribution, at ${\approx}3.5\sigma$ of its empirical width.
Any candidate with a secondary avalanche within $-40$ to $+140$ samples of the CFD reference time is removed, so that pileup does not enter the template.

\subsection{Signal Model and Analytic Marginalization}
\label{subsec:pileup-model}

The fit models each waveform as a sum of $K$ single-photon templates and searches for the number of photons and their arrival times that best explain it.
Integrating out the photon amplitudes analytically leaves a search over the count and the arrival times alone.
We optimize over an unbinned, continuous coordinate vector $\boldsymbol{\tau} = [\tau_1, \tau_2, \dots, \tau_K]^{\top} \in \mathbb{R}^K$, where the photon number $K$ acts as a trans-dimensional fluctuating variable.
Continuous sub-sample shifts are applied to the calibrated single photon template spectrum by complex phase rotations:
\begin{equation}
\widehat{Y}_{\mathrm{model}}(\omega) = \widehat{\mathcal{T}}(\omega) \sum_{k=1}^K A_k\, e^{-i \omega \tau_k}.
\end{equation}
The empirical template ($M$ samples) is much shorter than the observation window ($N$ samples), so shifting in the frequency domain would wrap late-arriving photons around to the start of the record.
To enforce a linear convolution, we evaluate the zero-padded superpositions in an extended $L$-dimensional space ($L \ge N + M - 1$).
The inverse FFT transforms them back to the time domain, truncating them at the observation boundary $N$.

To resolve the degeneracy of a singular matrix introduced by perfectly overlapping templates (as produced by prompt optical crosstalk), we impose a Tikhonov prior~\cite{Tikhonov:1963} on the amplitude vector $\mathbf{A}$, centered at the empirical single photon expectation $\boldsymbol{\mu}_{1\mathrm{pe}}$.
The unmarginalized scalar energy $E(\mathbf{A}, \boldsymbol{\tau})$, evaluated with the exact colored-noise covariance $\boldsymbol{\Sigma}$, is
\begin{equation}
E(\mathbf{A}, \boldsymbol{\tau}) \equiv \frac{1}{2} \|\mathbf{y} - \mathbf{Y}_{\mathrm{model}}\|^2_{\boldsymbol{\Sigma}^{-1}} + \frac{1}{2} \|\mathbf{A} - \boldsymbol{\mu}_{1\mathrm{pe}}\|^2_{\boldsymbol{\Lambda}^{-1}},
\end{equation}
where $\boldsymbol{\Lambda} = \sigma_A^2 \mathbf{I}_K$ is the isotropic gain covariance.
The model is linear in $\mathbf{A}$, so the amplitudes are marginalized analytically.
Published photomultiplier waveform analyses~\cite{Xu:2021whl,Wang:2024qxq} use the same construction, with the amplitudes integrated out analytically and, in FSMP~\cite{Wang:2024qxq}, a Gaussian single photon charge distribution as the prior.
The maximum a posteriori (MAP) amplitude estimate $\hat{\mathbf{A}}$ solves the normal equations at each step
\begin{equation}
\hat{\mathbf{A}} = \left( \mathbf{T}_{\boldsymbol{\tau}}^{\top} \boldsymbol{\Sigma}^{-1} \mathbf{T}_{\boldsymbol{\tau}} + \boldsymbol{\Lambda}^{-1} \right)^{-1} \left( \mathbf{T}_{\boldsymbol{\tau}}^{\top} \boldsymbol{\Sigma}^{-1} \mathbf{y} + \boldsymbol{\Lambda}^{-1} \boldsymbol{\mu}_{1\mathrm{pe}} \right),
\end{equation}
where $\mathbf{T}_{\boldsymbol{\tau}}$ is the $N \times K$ design matrix of unit templates at the proposed arrival times.

After substituting $\hat{\mathbf{A}}$ and integrating over the amplitudes, the search space collapses to the temporal dimensions $\boldsymbol{\tau}$.
The dimension $K$ is given a geometric prior $P(K) \propto e^{-K Z^{2}/2}$, equivalently a minimum significance per added photon.
The final marginalized energy $E_{\mathrm{marg}}(K, \boldsymbol{\tau})$, minimized by the search, becomes
\begin{equation}
\begin{split}
    E_{\mathrm{marg}}(K, \boldsymbol{\tau}) = {} &
\frac{1}{2} \| \mathbf{y} - \mathbf{Y}_{\mathrm{model}} \|^2_{\boldsymbol{\Sigma}^{-1}}
    + \frac{1}{2} \|\hat{\mathbf{A}} - \boldsymbol{\mu}_{1\mathrm{pe}}\|^2_{\boldsymbol{\Lambda}^{-1}} \\
    & + \frac{1}{2}\ln \det \Big( \sigma_A^2\,
\mathbf{T}_{\boldsymbol{\tau}}^{\top} \boldsymbol{\Sigma}^{-1} \mathbf{T}_{\boldsymbol{\tau}} + \mathbf{I}_K \Big) + K \frac{Z^2}{2},
\end{split}
\label{eq:emarg}
\end{equation}
where the third term is the Occam factor of Bayesian model comparison, which penalizes overlapping templates whose amplitudes the data cannot separate.
$Z$ sets the threshold, in units of standard deviations, that a proposed photon must exceed to justify its added dimension.

\subsection{Exact Energy Evaluation by Precomputed Correlation Tables}

Every quantity in $E_{\mathrm{marg}}$ is a whitened inner product between pairs of three objects: the data $\mathbf{y}$, the template $\mathcal{T}$ shifted to the proposed times, and the template derivative $\mathcal{T}'$.
A proposal is a candidate configuration $(K, \boldsymbol{\tau})$ that the search puts forward for scoring.
The fit does not recompute shifted templates for each proposal.
The whitened auto- and cross-correlation tables $R_{\mathcal{T}\mathcal{T}}(\Delta)$ and $R_{\mathcal{T}'\mathcal{T}}(\Delta)$ and the matched filter stream $R_{\mathbf{y}\mathcal{T}}(\tau)$ are precomputed.
This construction is performed once per channel, before any waveform is fitted, at time offsets on a grid twenty times finer than the \SI{1}{\nano\second} sampling.
A proposed configuration $(K, \boldsymbol{\tau})$ is then scored from the tables alone.
The $K \times K$ normal-equation matrix of Eq.~\eqref{eq:emarg} comes from $R_{\mathcal{T}\mathcal{T}}$ at the pairwise lags, the data projections from $R_{\mathbf{y}\mathcal{T}}$, and the exact gradient $\partial E_{\mathrm{marg}} / \partial \boldsymbol{\tau}$ from $R_{\mathcal{T}'\mathcal{T}}$.
No per-proposal transform, interpolation, or filtering is applied to the waveform.
The evaluation of the energy, its gradient, and the small-$K$ Cholesky solve is combined in one compiled routine, so a single call scores a proposal.
Sub-grid shifts use the analytic Fourier phase rotation of \ref{subsec:pileup-model} only when the tables are constructed, preserving a per-proposal cost independent of the record length.

\subsection{Seeding, Moves, Refinement, and Consensus}

The RJMCMC search is seeded by a greedy matching pursuit algorithm~\cite{MallatZhang:1993} over a continuous dictionary of shifted templates.
The seed algorithm iteratively evaluates the noise-weighted matched filter of the residual waveform in the frequency domain, records the largest peak on the sample grid, and subtracts the shifted template analytically by a complex phase rotation.
This yields a high-quality initial arrival time vector $\boldsymbol{\tau}_{\mathrm{seed}}$.

Each chain alternates birth and death proposals with shift moves at fixed dimension.
A birth draws its arrival time from the matched filter residual stream and refines it at once to its local MAP value before the accept/reject decision, so that dimension jumps land at local minima rather than at raw seed positions.
A death removes the weakest photon.
These optimized proposals are deliberately not balanced by the reverse-move bookkeeping of exact trans-dimensional posterior sampling~\cite{Green:1995,Wang:2024qxq}.
The chain is run as an annealed search whose Metropolis rule constrains the descent, not as a sampler with the posterior as its invariant distribution.
Between jumps, all arrival times are refined one at a time toward the MAP solution with the exact gradient from the tables; the refinement converges in a few sweeps because, once the amplitudes are marginalized (\ref{subsec:pileup-model}), the arrival times are only weakly coupled to one another.
For heavily collinear pileup, an optional split-merge pair replaces one arrival by two adjacent arrivals (and the reverse).
The generalized-least-squares solve marginalizes the amplitudes again automatically, so the move needs no amplitude bookkeeping.
Each waveform is fit by two such chains.
The final count is the $K$ on which the chains agree, and the solution is the lowest-energy one among the agreeing chains.
Consensus suppresses the residual risk of a single chain freezing in a local trans-dimensional mode.

\subsection{Priors and Thresholds}

The fit has two scalar settings.
The birth threshold $Z$ of Eq.~\eqref{eq:emarg}, which also sets the geometric prior on $K$, is the evidence in standard deviations that a proposed photon must provide to justify its dimension.
It controls the model complexity and fixes the false-positive rate on empty waveforms; no other prior is placed on $K$.
The width $\sigma_A$ of the amplitude prior is a regularization set below the per-pulse spread measured in the Section~\ref{subsec:spe} calibration.
By design, the fit has no prior on the arrival times.
The published samplers~\cite{Xu:2021whl,Wang:2024qxq} place a scintillation light-curve prior on each photon time, which FSMP also uses to speed convergence~\cite{Wang:2024qxq}, whereas the emission-time response is a measurement target of this paper (Section~\ref{subsec:res-timing}).
An assumed profile included in the fit would bias the measurement toward it.
The closure of \ref{app:rjmcmc-closure} uses these production settings; its confusion matrix and timing residuals describe the production configuration, not a separately tuned benchmark.

\subsection{Closure Methodology}
\label{app:rjmcmc-closure}

\begin{figure}[htbp]
    \centering
    \includegraphics[width=\columnwidth]{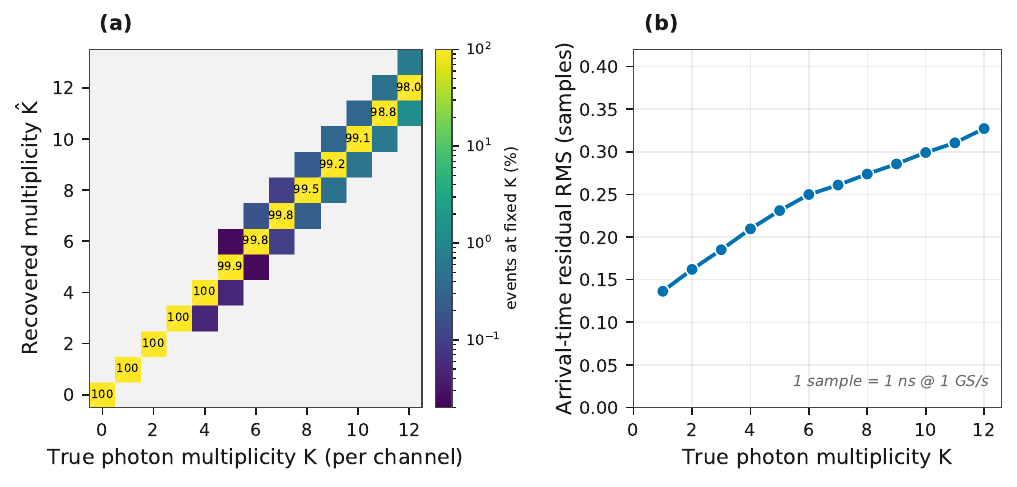}
\caption{Matched closure of the photon-counting fit: 4000 synthetic waveforms per multiplicity, generated from the single photon template and noise model of the fit itself and fitted with the production configuration.
(a) Confusion matrix of reconstructed versus true per-channel multiplicity, row-normalized at fixed $K$ (diagonal annotated in percent).
(b) Root-mean-square arrival time residual of matched photons as a function of $K$, in samples.}
    \label{fig:closure}
\end{figure}

Every downstream observable assumes that the fit reconstructs the counts $K$ and arrival times $\boldsymbol{\tau}$ from piled-up waveforms; a matched closure study checks that it does.
The synthetic waveforms are built from the \textit{in situ} single photon template and noise model of Section~\ref{subsec:spe} that the fit itself uses, with arrival times drawn from the analytic emission model.
They are processed through the production reconstruction chain and compared with the known truth.
Since the synthesis uses the same template and noise model as the fit, the test is self-consistent.
Any residual bias on real data must then come from effects outside the linear superposition model, such as nonlinearity of the digitizer electronics under dense pileup.
The fit returns the true count for all 4000 single-photoelectron waveforms, for $99.80\%$ of the $K=6$ waveforms, and for $98.0\%$ of the $K=12$ waveforms.
Miscounts are confined to $\hat{K} = K \pm 1$ (largest off-diagonal leakage $1.2\%$).
The false-positive rate on empty waveforms is $<0.025\%$, and matched arrival time residuals rise from $0.14$ to $0.33$ samples across $K = 1$--$12$ (Fig.~\ref{fig:closure}).
Synthetic waveforms drawn at the measured per-pulse amplitude spread and fitted with the deployed prior width, rather than with a matched width, differ by less than $0.03$ in the exact-multiplicity fraction at $K = 12$ and by less than $0.3\%$ in the mean count.

\subsection{Computational Profile}
\label{app:profile}

With the precomputed tables and the single routine, a typical selected waveform is fitted in ${\approx}\SI{6}{\milli\second}$ on one core, roughly forty times faster than the initial implementation with one transform per proposal.
The closure metrics of Fig.~\ref{fig:closure} are unchanged, the accuracy gates (exact-multiplicity fraction, matched-photon efficiency, false-positive rate, and arrival time residual) having been checked again at each optimization stage.
This throughput makes two analyses in the paper practical.
The full $1.3 \times 10^{6}$-trigger nominal run is processed end to end in under an hour on a single core.
Even the dark-noise campaign of Section~\ref{subsec:res-spe}, which fits every channel of every trigger with no selection, runs on a laptop.
Waveforms with no seed above threshold bypass the fit and are given $K = 0$, at the false-positive rate of $<0.025\%$ that the closure bounds (\ref{app:rjmcmc-closure}).

\section{Selection and Background Derivations}
\label{app:selection-derivations}

This appendix derives two quantities used in Sections~\ref{sec:results} and~\ref{sec:systematics}: the selection bias of the fiducial cut and the dark counts in the time window.

\subsection{Selection Bias of the Fiducial Cut}
\label{app:self-selection}
A cut on the reconstructed vertex passes an event with few photons, such as those at the lowest operating point (\SI{181.9}{keV}), more often when its photon count fluctuates upward, because the extra photons localize its vertex better.
This biases the accepted mean high.
Binomial photon thinning turns a bright run into a sample statistically identical to a fainter run at the same vertices.
Applying the cut again to the replicas measures the bias directly.
The bias at the lowest operating point is $+14$--$16\%$.
It falls roughly as one over the count.
This bias shifts only the fitted scale of the mixture comparison in Section~\ref{subsec:res-resolution}.
Refitting all four points at a common thinned light level shifts every point by the same factor and leaves the slope unchanged.
The per-MeV yield ratios between the $^{137}$Cs and $^{60}$Co points at each angle, used for the quenching test of Section~\ref{subsec:res-lightyield}, are computed on these replicas with the background inside the coincidence window subtracted.
The slope without the vertex requirement is $47.2$~p.e.\ per MeV in counted avalanches, compared to the fitted $48.3 \pm 1.5$~p.e.\ per MeV\@.
A scan of the cut radius from 3 to \SI{15}{\milli\meter} changes the central yield by less than $\pm 1\%$ with no trend.

\subsection{The Dark Counts in the Time Window}
\label{app:dark-floor}
Sixty live channels at the ${\approx}\SI{160}{\kilo\hertz}$ mean per-channel rate give $60 \times \SI{160}{\kilo\hertz} \times \SI{105}{\nano\second} \approx 1.0$ counted avalanches per event across the \SI{105}{\nano\second} time window, independent of deposit energy.
This count enters the linearity intercept of Section~\ref{subsec:res-lightyield} and the Poisson mixture of Section~\ref{subsec:res-resolution}.

\bibliographystyle{elsarticle-num}
\bibliography{bibliography}

\begin{thebibliography}{10}
\expandafter\ifx\csname url\endcsname\relax
  \def\url#1{\texttt{#1}}\fi
\expandafter\ifx\csname urlprefix\endcsname\relax\def\urlprefix{URL }\fi
\expandafter\ifx\csname href\endcsname\relax
  \def\href#1#2{#2} \def\path#1{#1}\fi

\bibitem{NavasNicolas:2024ixn}
D.~Navas-Nicol{\'a}s, {CLOUD: the First Reactor Antineutrino Experiment using
  the Novel LiquidO Detection Technology}, PoS ICHEP2024 (2025) 157.
\newblock \href {https://doi.org/10.22323/1.476.0157}
  {\path{doi:10.22323/1.476.0157}}.

\bibitem{NuDoubt:2024jax}
M.~B{\"o}hles, et~al., {Combining hybrid and opaque scintillator techniques in
  the search for double beta plus decays}, Eur. Phys. J. C 85~(2) (2025) 121,
  [Erratum: Eur. Phys. J. C 85, 755 (2025)].
\newblock \href {http://arxiv.org/abs/2407.05999} {\path{arXiv:2407.05999}},
  \href {https://doi.org/10.1140/epjc/s10052-025-13847-1}
  {\path{doi:10.1140/epjc/s10052-025-13847-1}}.

\bibitem{LiquidO:2025hbe}
S.~R. Soleti, et~al., {COCOA: A compact Compton camera for astrophysical
  observation of MeV-scale gamma rays}, Astropart. Phys. 172 (2025) 103135.
\newblock \href {http://arxiv.org/abs/2502.20916} {\path{arXiv:2502.20916}},
  \href {https://doi.org/10.1016/j.astropartphys.2025.103135}
  {\path{doi:10.1016/j.astropartphys.2025.103135}}.

\bibitem{LiquidO:2019mxd}
A.~Cabrera, et~al., {Neutrino Physics with an Opaque Detector}, Commun. Phys. 4
  (2021) 273.
\newblock \href {http://arxiv.org/abs/1908.02859} {\path{arXiv:1908.02859}},
  \href {https://doi.org/10.1038/s42005-021-00763-5}
  {\path{doi:10.1038/s42005-021-00763-5}}.

\bibitem{Wilhelm:2023hty}
A.~S. Wilhelm, G.~Wendel, B.~Collins, D.~Cowen, I.~Jovanovic, {Evaluation of
  light collection from highly scattering media using wavelength-shifting
  fibers}, Nucl. Instrum. Meth. A 1049 (2023) 168085.
\newblock \href {http://arxiv.org/abs/2301.09608} {\path{arXiv:2301.09608}},
  \href {https://doi.org/10.1016/j.nima.2023.168085}
  {\path{doi:10.1016/j.nima.2023.168085}}.

\bibitem{LiquidO:2025qia}
J.~Apilluelo, et~al., {The Stochastic Light Confinement of LiquidO} (2025).
\newblock \href {http://arxiv.org/abs/2503.02541} {\path{arXiv:2503.02541}}.

\bibitem{LiquidO:2024piw}
J.~Apilluelo, et~al., {Characterization of a radiation detector based on opaque
  water-based liquid scintillator}, Nucl. Instrum. Meth. A 1071 (2025) 170075.
\newblock \href {http://arxiv.org/abs/2406.13054} {\path{arXiv:2406.13054}},
  \href {https://doi.org/10.1016/j.nima.2024.170075}
  {\path{doi:10.1016/j.nima.2024.170075}}.

\bibitem{LiquidO:2025rhe}
J.~Apilluelo, et~al., {Muon tracking in a LiquidO opaque scintillator
  detector}, JINST 21~(01) (2026) P01010.
\newblock \href {http://arxiv.org/abs/2507.13864} {\path{arXiv:2507.13864}},
  \href {https://doi.org/10.1088/1748-0221/21/01/P01010}
  {\path{doi:10.1088/1748-0221/21/01/P01010}}.

\bibitem{Che:2026wab}
H.~Che, M.~V. Diwan, S.~Gokhale, P.~Kumar, C.~Reyes, R.~Rosero, J.~J. Wang,
  G.~Yang, M.~Yeh, {Construction, commissioning, and beam test of a pilot
  3D-projection opaque water-based liquid scintillator detector} (2026).
\newblock \href {http://arxiv.org/abs/2605.10683} {\path{arXiv:2605.10683}}.

\bibitem{Buck:2019tsa}
C.~Buck, B.~Gramlich, S.~Schoppmann, {Novel Opaque Scintillator for Neutrino
  Detection}, JINST 14~(11) (2019) P11007.
\newblock \href {http://arxiv.org/abs/1908.03334} {\path{arXiv:1908.03334}},
  \href {https://doi.org/10.1088/1748-0221/14/11/P11007}
  {\path{doi:10.1088/1748-0221/14/11/P11007}}.

\bibitem{Beretta:2025jinst}
M.~Beretta, et~al., {Fluorescence emission of the JUNO liquid scintillator},
  JINST 20~(05) (2025) P05009.
\newblock \href {http://arxiv.org/abs/2501.09988} {\path{arXiv:2501.09988}},
  \href {https://doi.org/10.1088/1748-0221/20/05/P05009}
  {\path{doi:10.1088/1748-0221/20/05/P05009}}.

\bibitem{Ribberfors:1975}
R.~Ribberfors, {Relationship of the relativistic Compton cross section to the
  momentum distribution of bound electron states}, Phys. Rev. B 12 (1975)
  2067--2074.
\newblock \href {https://doi.org/10.1103/PhysRevB.12.2067}
  {\path{doi:10.1103/PhysRevB.12.2067}}.

\bibitem{KleinNishina:1929}
O.~Klein, Y.~Nishina, {\"{U}ber die Streuung von Strahlung durch freie
  Elektronen nach der neuen relativistischen Quantendynamik von Dirac}, Z.
  Phys. 52 (1929) 853--868.
\newblock \href {https://doi.org/10.1007/BF01366453}
  {\path{doi:10.1007/BF01366453}}.

\bibitem{Macchiarulo:2023hdsocv2}
L.~Macchiarulo, I.~Mostafanezhad, G.~S. Varner, C.~White, D.~Uehara, R.~Perron,
  G.~Uehara, C.~Chock, {Design and Initial Measurement Results for the second
  revision of the High Density Digitizer System on Chip (HDSoCv2): A 64 Channel
  1 GSa/s Waveform Digitizer for High Density Detectors}, in: 2023 IEEE Nuclear
  Science Symposium, Medical Imaging Conference and International Symposium on
  Room-Temperature Semiconductor Detectors (NSS MIC RTSD), IEEE, 2023.
\newblock \href {https://doi.org/10.1109/nssmicrtsd49126.2023.10338148}
  {\path{doi:10.1109/nssmicrtsd49126.2023.10338148}}.

\bibitem{Macchiarulo:2024hdsocv2}
L.~Macchiarulo, I.~Mostafanezhad, K.~Hoe, B.~Rotter, D.~Uehara, C.~Chock,
  {Measurements of the High Density Digitizer System on Chip (HDSoCv2): A 64
  Channel 1 GSa/s Waveform Digitizer for High Density Detectors}, in: 2024 IEEE
  Nuclear Science Symposium (NSS), Medical Imaging Conference (MIC) and Room
  Temperature Semiconductor Detector Conference (RTSD), IEEE, 2024.
\newblock \href {https://doi.org/10.1109/nss/mic/rtsd57108.2024.10656600}
  {\path{doi:10.1109/nss/mic/rtsd57108.2024.10656600}}.

\bibitem{GEANT4:2002zbu}
S.~Agostinelli, et~al., {GEANT4--a simulation toolkit}, Nucl. Instrum. Meth. A
  506 (2003) 250--303.
\newblock \href {https://doi.org/10.1016/S0168-9002(03)01368-8}
  {\path{doi:10.1016/S0168-9002(03)01368-8}}.

\bibitem{Allison:2006ve}
J.~Allison, et~al., {Geant4 developments and applications}, IEEE Trans. Nucl.
  Sci. 53 (2006) 270--278.
\newblock \href {https://doi.org/10.1109/TNS.2006.869826}
  {\path{doi:10.1109/TNS.2006.869826}}.

\bibitem{Allison:2016lfl}
J.~Allison, et~al., {Recent developments in Geant4}, Nucl. Instrum. Meth. A 835
  (2016) 186--225.
\newblock \href {https://doi.org/10.1016/j.nima.2016.06.125}
  {\path{doi:10.1016/j.nima.2016.06.125}}.

\bibitem{ratpac-two}
{RAT-PAC~2 developers}, {RAT-PAC~2: RAT Plus Additional Codes, version 2},
  \url{https://github.com/rat-pac/ratpac-two}, version 3.4.0 (commit 6827472)
  with Geant4 11.3.p02 for the gamma-ray transport; version 3.3.0 with Geant4
  11.1.p02 for the optical transport (2026).

\bibitem{Herzkamp2016SimulationNeutrons}
M.~Herzkamp, {Simulation and Optimization of a Position Sensitive Scintillation
  Detector with Wavelength Shifting Fibers for Thermal Neutrons}, Ph.D. thesis,
  RWTH Aachen University (2016).

\bibitem{Levin:1996}
A.~Levin, C.~Moisan, {A more physical approach to model the surface treatment
  of scintillation counters and its implementation into DETECT}, in: 1996 IEEE
  Nuclear Science Symposium. Conference Record, Vol.~2, 1996, pp. 702--706.
\newblock \href {https://doi.org/10.1109/NSSMIC.1996.591410}
  {\path{doi:10.1109/NSSMIC.1996.591410}}.

\bibitem{Brent:1973}
R.~P. Brent, {Algorithms for Minimization without Derivatives}, Prentice-Hall,
  Englewood Cliffs, NJ, 1973.

\bibitem{Green:1995}
P.~J. Green, {Reversible jump Markov chain Monte Carlo computation and Bayesian
  model determination}, Biometrika 82~(4) (1995) 711--732.
\newblock \href {https://doi.org/10.1093/biomet/82.4.711}
  {\path{doi:10.1093/biomet/82.4.711}}.

\bibitem{Xu:2021whl}
D.~Xu, et~al., {Towards the ultimate PMT waveform analysis for neutrino and
  dark matter experiments}, JINST 17~(06) (2022) P06040.
\newblock \href {http://arxiv.org/abs/2112.06913} {\path{arXiv:2112.06913}},
  \href {https://doi.org/10.1088/1748-0221/17/06/P06040}
  {\path{doi:10.1088/1748-0221/17/06/P06040}}.

\bibitem{Wang:2024qxq}
Y.~Wang, A.~Zhang, Y.~Wu, B.~Xu, X.~Liu, J.~Chen, Z.~Wang, S.~Chen, {The Fast
  Stochastic Matching Pursuit for neutrino and dark matter experiments}, Nucl.
  Instrum. Meth. A 1082 (2026) 170986.
\newblock \href {http://arxiv.org/abs/2403.03156} {\path{arXiv:2403.03156}},
  \href {https://doi.org/10.1016/j.nima.2025.170986}
  {\path{doi:10.1016/j.nima.2025.170986}}.

\bibitem{NepomukOtte:2016ktf}
A.~N. Otte, D.~Garcia, T.~Nguyen, D.~Purushotham, {Characterization of Three
  High Efficiency and Blue Sensitive Silicon Photomultipliers}, Nucl. Instrum.
  Meth. A 846 (2017) 106--125.
\newblock \href {http://arxiv.org/abs/1606.05186} {\path{arXiv:1606.05186}},
  \href {https://doi.org/10.1016/j.nima.2016.09.053}
  {\path{doi:10.1016/j.nima.2016.09.053}}.

\bibitem{JUNO:2024fdc}
A.~Abusleme, et~al., {Prediction of Energy Resolution in the JUNO Experiment},
  Chin. Phys. C 49~(1) (2025) 013003.
\newblock \href {http://arxiv.org/abs/2405.17860} {\path{arXiv:2405.17860}},
  \href {https://doi.org/10.1088/1674-1137/ad83aa}
  {\path{doi:10.1088/1674-1137/ad83aa}}.

\bibitem{DayaBay:2019fje}
D.~Adey, et~al., {A high precision calibration of the nonlinear energy response
  at Daya Bay}, Nucl. Instrum. Meth. A 940 (2019) 230--242.
\newblock \href {http://arxiv.org/abs/1902.08241} {\path{arXiv:1902.08241}},
  \href {https://doi.org/10.1016/j.nima.2019.06.031}
  {\path{doi:10.1016/j.nima.2019.06.031}}.

\bibitem{Birks:1951}
J.~B. Birks, {Scintillations from organic crystals: specific fluorescence and
  relative response to different radiations}, Proc. Phys. Soc. A 64 (1951)
  874--877.
\newblock \href {https://doi.org/10.1088/0370-1298/64/10/303}
  {\path{doi:10.1088/0370-1298/64/10/303}}.

\bibitem{Kodama:2023suc}
S.~Kodama, H.~Kobayashi, W.~Okinaga, K.~Nakagiri, Y.~Nakajima, M.~Yokoyama,
  {Performance of New Kuraray Wavelength-shifting Fibers with Short Decay
  Time}, PTEP 2024~(5) (2024) 053H01.
\newblock \href {http://arxiv.org/abs/2311.07297} {\path{arXiv:2311.07297}},
  \href {https://doi.org/10.1093/ptep/ptae055}
  {\path{doi:10.1093/ptep/ptae055}}.

\bibitem{Seifert:2012pmb}
S.~Seifert, H.~T. van Dam, D.~R. Schaart, {The lower bound on the timing
  resolution of scintillation detectors}, Phys. Med. Biol. 57~(7) (2012)
  1797--1814.
\newblock \href {https://doi.org/10.1088/0031-9155/57/7/1797}
  {\path{doi:10.1088/0031-9155/57/7/1797}}.

\bibitem{van_Loef_2001}
E.~V.~D. van Loef, P.~Dorenbos, C.~W.~E. van Eijk, K.~Kr{\"a}mer, H.~U.
  G{\"u}del, {High-energy-resolution scintillator: Ce$^{3+}$ activated
  LaBr$_3$}, Appl. Phys. Lett. 79~(10) (2001) 1573--1575.
\newblock \href {https://doi.org/10.1063/1.1385342}
  {\path{doi:10.1063/1.1385342}}.

\bibitem{Moszynski:2002kr}
M.~Moszynski, J.~Zalipska, M.~Balcerzyk, M.~Kapusta, W.~Mengesha, J.~D.
  Valentine, {Intrinsic energy resolution of NaI(Tl)}, Nucl. Instrum. Meth. A
  484 (2002) 259--269.
\newblock \href {https://doi.org/10.1016/S0168-9002(01)01964-7}
  {\path{doi:10.1016/S0168-9002(01)01964-7}}.

\bibitem{Glodo_2005}
J.~Glodo, W.~W. Moses, W.~M. Higgins, E.~V.~D. van Loef, P.~Wong, S.~E.
  Derenzo, M.~J. Weber, K.~S. Shah, {Effects of Ce concentration on
  scintillation properties of LaBr$_3$:Ce}, IEEE Trans. Nucl. Sci. 52~(5)
  (2005) 1805--1808.
\newblock \href {https://doi.org/10.1109/TNS.2005.856906}
  {\path{doi:10.1109/TNS.2005.856906}}.

\bibitem{Tikhonov:1963}
A.~N. Tikhonov, {Solution of incorrectly formulated problems and the
  regularization method}, Sov. Math. Dokl. 4 (1963) 1035--1038.

\bibitem{MallatZhang:1993}
S.~G. Mallat, Z.~Zhang, {Matching pursuits with time-frequency dictionaries},
  IEEE Trans. Signal Process. 41~(12) (1993) 3397--3415.
\newblock \href {https://doi.org/10.1109/78.258082}
  {\path{doi:10.1109/78.258082}}.

\end{thebibliography}

\typeout{get arXiv to do 4 passes: Label(s) may have changed. Rerun}
\end{document}